\documentclass[a4paper,fleqn]{cas-sc}

\usepackage[numbers]{natbib}
\usepackage{bm}
\usepackage{setspace}
\usepackage{lineno}
\newcommand{\bi}[1]{\textbf{\textit{#1}}}

\def\tsc#1{\csdef{#1}{\textsc{\lowercase{#1}}\xspace}}
\tsc{WGM}
\tsc{QE}
\tsc{EP}
\tsc{PMS}
\tsc{BEC}
\tsc{DE}
\begin{document}
\let\WriteBookmarks\relax
\def\floatpagepagefraction{1}
\def\textpagefraction{.001}
\shorttitle{Improving Observability of Relative Orbit Estimation Using Bearing Measurements and Light Curves}
\shortauthors{Yasuhiro Yoshimura et~al.}
%\begin{frontmatter}

\title [mode = title]{Improving Observability of Relative Orbit Estimation Using Bearing Measurements and Light Curves}
% \tnotemark[1,2]

% \tnotetext[1]{This document is the results of the research
% project funded by the National Science Foundation.}

% \tnotetext[2]{The second title footnote which is a longer text matter
% to fill through the whole text width and overflow into
% another line in the footnotes area of the first page.}

\author[1]{Yasuhiro Yoshimura}[type=editor,
     auid=000,bioid=1]
\cormark[1]
\fnmark[1]
\ead{y.yoshimura.a64@m.kyushu-u.ac.jp}
\credit{Conceptualization of this study, Methodology, Software, Writing - Original draft preparation}

\affiliation[1]{organization={Department of Aeronautics and Astronautics, Kyushu University},
     addressline={744 Motooka, Nishi-ku},
     city={Fukuoka},
     postcode={819-0395},
     state={Fukuoka},
     country={Japan}}

\author[2]{Toshiya Hanada}[auid=000,bioid=2]
\fnmark[2]
\credit{Data curation, Supervision}

% \cormark[2]
% \fnmark[1,3]
% \ead{t.rafeeq@example.in}
% \ead[URL]{www.campus.in}

\cortext[cor1]{Corresponding author}
% \cortext[cor2]{Principal corresponding author}
% \fntext[fn1]{This is the first author footnote, but is common to third
% \fntext[fn2]{Another author footnote, this is a very long footnote and
%   it should be a really long footnote. But this footnote is not yet
%   sufficiently long enough to make two lines of footnote text.}

% \nonumnote{This note has no numbers. In this work we demonstrate $a_b$
%   the formation Y\_1 of a new type of polariton on the interface
%   between a cuprous oxide slab and a polystyrene micro-sphere placed

\begin{abstract}
     Relative orbit estimation using optical observations is a key technology for on-orbit servicing missions. In the far-range phase, the target appears as an unresolved point source, providing only bearing angles (azimuth and elevation) from the servicing satellite. Angles-only navigation is inherently challenging due to the weak observability of the relative range. To address this limitation, this study investigates the effectiveness of an estimation scheme that fuses photometric light curve data with bearing measurements. Since the light intensity depends on the relative distance, fusing light curves enhances the observability of the relative state. The Ashikhmin-Shirley model is used as the optical reflectance model, and observability analysis is conducted with the Fisher information matrix. Numerical simulations involving different target geometries, a flat plate and a box-wing satellite, demonstrate that integrating light curve measurements significantly enhances observability and enables faster convergence compared to conventional state estimation methods.

\end{abstract}

\begin{graphicalabstract}
\end{graphicalabstract}

\begin{highlights}
     \item Observability analysis with bearing and light curve measurements is conducted.
     \item Integrating light curve measurements enables faster convergence of the relative orbit estimation.
     \item The estimation scheme is robust against large initial relative orbit errors.
\end{highlights}

\begin{keywords}
     Relative orbit estimation \sep Light curves \sep Observability analysis \sep On-orbit servicing \sep Formation flying \\
\end{keywords}

\maketitle
% doublespacing removed for arXiv (single-spaced required)
% linenumbers removed for arXiv

\section{Introduction}
On-orbit servicing missions~\cite{dremannReviewAdvancementsInspection2025}, including refueling~\cite{hattyViabilityOnOrbitServicing2022, alandihallajModelPredictiveControlbased2025}, inspection~\cite{nakkaInformationBasedGuidanceControl2022,nobuharaOptimalTrajectoryDesign2026}, and active debris removal~\cite{svotinaSpaceDebrisRemoval2024}, necessitate autonomous capabilities for target acquisition and proximity operations.
In the far-range phase, the servicing satellite observes the target using optical cameras.
The target typically appears as an unresolved point source, yielding a two-dimensional position on the image plane.
These measurements correspond to the azimuth and elevation angles (bearing angles) from the servicing satellite to the target.
Relative orbit estimation relying solely on these angles is known as angles-only navigation~\cite{gongRBFNNbasedAnglesonlyOrbit2024,kruger2024}.
A primary challenge in angles-only navigation is the inherent weak observability of the relative range~\cite{Woffinden2009}.
The observability of nonlinear relative motion is analyzed in~\cite{butcher2017}, showing that the relative state is locally weakly observable.

To address this challenge, orbital maneuvers are used to enhance observability in~\cite{GaiasJGCD}, while the use of a camera offset is considered in~\cite{geller2014}.
Observability analysis incorporating orbital perturbations due to Earth's oblateness is examined in~\cite{Psiaki}.
A fuel-free estimation method is introduced in~\cite{Sullivan2017} and extended to formation flying with multiple spacecraft in~\cite{generalized}.
Hu et al.~\cite{Hu} examine three-spacecraft formation flying and its angles-only navigation, which is subsequently generalized to accommodate an arbitrary number of spacecraft in~\cite{Hu2}.
Given the significance of nonlinearity in estimating weakly observable distances, precise modeling of the line of sight is explored in~\citep{gaias2021}, while Givens and McMahon~\cite{givens} focus on spherical coordinates to facilitate a linear measurement model.
An overdetermined eigenvector approach is also proposed for passive angles-only relative orbit determination~\cite{Givens2024eigenvector}.
Regarding machine learning approaches, Gong et al.~\cite{gong} present an estimation method using neural networks based on radial basis functions, effectively capturing the nonlinearity inherent in angles-only navigation.
Recent work by Kruger and D'Amico~\cite{kruger2024} presents observability analysis and optimization for angles-only navigation of distributed space systems, and the Starling Formation-Flying Optical Experiment (StarFOX) demonstrates the first flight demonstration of angles-only navigation for a swarm of spacecraft~\cite{StarFOX2023}.
Stoker-Spirt et al.~\cite{stoker-spirt_aiding_2023} propose aiding angles-only navigation with illumination telemetry to resolve weak observability, which is closely related to the approach proposed in this paper.

In the context of improving observability, this paper investigates fusing bearing angle measurements with photometric observations of the unresolved target, known as light curves.
Since the measured light intensity depends on the target's shape, attitude, optical properties, and relative distance, combining light curves with bearing angles significantly enhances observability for relative orbit estimation.
Light curves have been used for estimating the state of resident space objects in the context of space situational awareness~\citep{Dianetti2023-gp, haraAttitudeEstimationPhotometric2025}.

The idea of aiding angles-only relative navigation with photometric brightness is itself not new. Whittaker et al.~\cite{whittaker2013} fuse bearing angles with photometric light intensity in an unscented Kalman filter to estimate the relative position and velocity of formation-flying spacecraft, mitigating the range ambiguity of angles-only navigation. That study, however, models the target as a simple diffuse sphere, assumes cooperative spacecraft, and provides an initial feasibility demonstration without an observability analysis. The obstacle that has kept photometric relative navigation at this proof-of-concept stage is the strong dependence of the measured luminosity on the unknown attitude and optical properties of a non-cooperative target. This difficulty is made explicit by the AVANTI flight demonstration of angles-only navigation to a non-cooperative target~\cite{ardaensGaias2018}, which recovers the weakly observable range direction through calibrated, propellant-consuming maneuvers and deliberately excludes apparent magnitude from its navigation filter, exploiting it only for target discrimination because it depends strongly on the unknown attitude and illumination conditions. The present study overcomes this obstacle by exploiting the fact that the range information carried by the light curve resides in a term that is independent of the absolute brightness level, and hence of the detailed shape and optical model. With the target attitude treated as a known time-varying input, this term is exposed and supplies, passively and without maneuvers, the range information that AVANTI obtained actively. This paper thus advances photometric relative navigation from the cooperative diffuse-sphere proof of concept toward a realistic, non-cooperative, observability-quantified treatment.

To accurately model light curves, physically based reflectance models, such as the Ashikhmin--Shirley model~\citep{ashikhmin2000}, are usually employed.
Since these reflectance models are highly nonlinear functions of the target state, an adaptive unscented Kalman filter (UKF) is adopted. The estimated state is augmented to include the diffuse reflectance of the target surface to enhance robustness against uncertainty in the optical properties.
Monte Carlo simulations involving two target geometries, a flat plate and a box-wing satellite, are conducted to investigate the effectiveness of fusing light curve measurements with bearing angle measurements.

\section{Preliminaries}
\subsection{Orbital Motion and Bearing Angles}
\begin{figure}[tb]
     \centering
     \includegraphics[width=7cm]{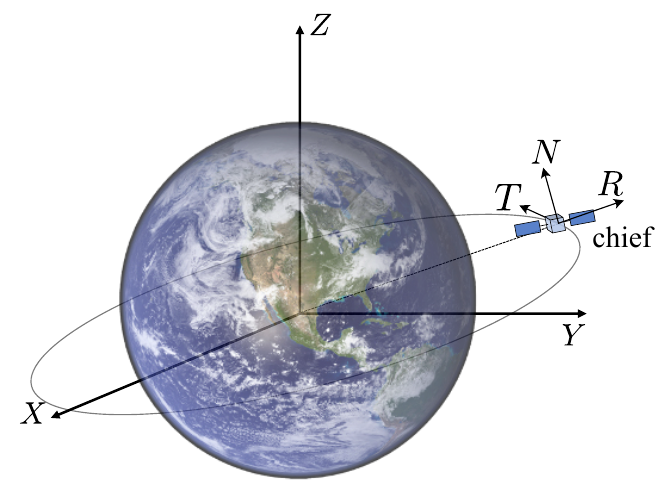}
     \caption{RTN frame and inertial frame.}
     \label{fig:rtnFrame}
\end{figure}

The relative motion of the deputy satellite (target object) with respect to the chief satellite is described in the Radial-Transversal-Normal (RTN) frame as shown in Fig.~\ref{fig:rtnFrame}.
The RTN frame has its origin at the center of mass of the chief; the $R$ axis points from the Earth's center to the chief, the $N$ axis is aligned with the chief's orbital angular momentum, and the $T$ axis completes the right-handed coordinate system.
The inertial frame has its origin at the center of mass of the Earth; the $X$ axis is aligned with the vernal equinox direction, the $Z$ axis is aligned with the Earth's rotation axis, and the $Y$ axis completes the right-handed coordinate system.
The body-fixed frames of the satellites are assumed to coincide with their principal axes of inertia.

This paper employs dimensionless quasi-nonsingular relative orbital elements (ROEs)~\citep{Sullivan2017} to represent the relative motion of the deputy satellite because they are slow-varying parameters.
The ROEs $\delta\bi{\oe}$ are defined as
\begin{align}
     \delta\bi{\oe} = \begin{bmatrix}
                           \delta a       \\
                           \delta \lambda \\
                           \delta e_{x}   \\
                           \delta e_{y}   \\
                           \delta i_{x}   \\
                           \delta i_{y}
                      \end{bmatrix} = \left[\begin{array}{c}
                                                      \frac{a_{d} - a}{a}                     \\
                                                      u_d - u + (\Omega_{d} - \Omega) \cos{i} \\
                                                      e_{x,d} - e_{x}                         \\
                                                      e_{y, d} - e_{y}                        \\
                                                      i_{d } - i                              \\
                                                      (\Omega_{d} - \Omega) \sin{i}
                                                 \end{array}\right] \label{eq:roeDef}
\end{align}
where $a$ is the semi-major axis, $u=w + M$ is the mean argument of latitude defined by the argument of perigee $w$ and the mean anomaly $M$, $\Omega$ is the right ascension of the ascending node, and $i$ is the inclination.
The subscript~$d$ denotes the orbital elements of the deputy satellite, while elements without a subscript refer to those of the chief.
The components of the eccentricity vector are expressed as $e_{x} = e\cos{w}$ and $e_{y} = e\sin{w}$.
A non-zero $\delta a$ induces drift motion, causing variations in relative distance and leading to a time-varying relative mean longitude $\delta \lambda$.
Consequently, the relative range information, which is consistently difficult to observe, is primarily contained within $\delta \lambda$.
Using the ROEs and the absolute orbital elements of the chief, the absolute orbital elements of the deputy can be derived from Eq.~\eqref{eq:roeDef} as
\begin{align}
     a_{d}                    & = a\delta a + a                                                         \\
     u_{d}                    & = u + \delta\lambda - (\Omega_{d} - \Omega)\cos{i}                      \\
     e_{d}                    & = \sqrt{(e_{x}+\delta e_{x})^{2} + (e_{y}+ \delta e_{y})^{2}}           \\
     i_{d}                    & = i + \delta i_{x}                                                      \\
     \displaystyle w_{d}      & = \arctan{\left(\frac{e_{y}+ \delta e_{y}}{e_{x}+ \delta e_{x}}\right)} \\
     \displaystyle \Omega_{d} & = \Omega + \frac{\delta i_{y}}{\sin{i}}
\end{align}

The relative position of the deputy in the RTN frame, $\delta \bm{r}^\mathcal{O}_{d}$, can be obtained using either a nonlinear or a linear method.
The nonlinear method calculates the absolute position and velocity of both satellites in the inertial frame, determining the relative position by taking their difference and transforming it to the RTN frame.
The linear method employs a mapping from ROEs to the RTN frame~\cite{gaias2015}, simplifying the calculation and reducing computational cost. The mapping is expressed as
\begin{align}
     \delta \bm{r}_{d}^{\mathcal{O}} & = a\frac{\partial \delta \bm{r}_{d}^{\mathcal{O}}}{\partial \delta \bi{\oe}} \delta \bi{\oe}
\end{align}
where
\begin{align}
     \frac{\partial \delta \boldsymbol{r}_{d}^{\mathcal{O}}}{\partial \delta \bi{\oe}} & = \begin{bmatrix}
                                                                                                1 & 0 & -\cos{u} & -\sin{u}  & 0       & 0        \\
                                                                                                0 & 1 & 2\sin{u} & -2\cos{u} & 0       & 0        \\
                                                                                                0 & 0 & 0        & 0         & \sin{u} & -\cos{u}
                                                                                           \end{bmatrix}
\end{align}
This linear mapping is used only in the observability analysis, whereas the full nonlinear transformation is used for the relative motion estimation. Within the observability analysis, the Fisher information matrix constitutes a first-order, local Cram\'er--Rao bound, for which this linearized mapping is exactly the required Jacobian rather than an avoidable approximation. The neglected terms are second order in the relative orbital element magnitude and are therefore small for the near-circular, small-separation regime considered in this paper. Moreover, the observability index is used only to contrast the angles-only and angles with light curve cases, and any residual linearization bias is common to both and cancels in the comparison.

\begin{figure}[tb]
     \centering
     \includegraphics[width=7cm]{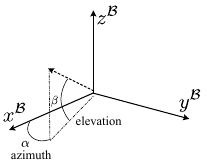}
     \caption{Definition of azimuth and elevation angles.}
     \label{fig:aziEle}
\end{figure}

Assuming the chief's optical camera is fixed to the satellite body and the camera-fixed frame is aligned with the body-fixed frame, the azimuth and elevation angles $(\alpha, \beta)$ are defined as (see also Fig.~\ref{fig:aziEle}):
\begin{align}
     \alpha & = \arctan{\left(\frac{y}{x}\right)} \label{eq:azi}                                   \\
     \beta  & = \arcsin{\left(\frac{z}{\|\delta \bm{r}_{d}^{\mathcal{B}}\|}\right)} \label{eq:ele}
\end{align}
where $\delta \bm{r}_{d}^{\mathcal{B}} = [x,y,z]^{T}$ is the relative position of the deputy in the chief's body-fixed frame.

The time derivatives of the absolute orbital elements under a perturbing acceleration are given by the Gauss variational equations (GVE) as
\begin{align}
     \frac{\mathrm{d}a}{\mathrm{d}t}      & = \frac{2}{n_{o}\sqrt{1 - e^{2}}}\left[e\sin{(f)}a_{R} + (1+e\cos{f})a_{T}\right]  \label{eq:GaussVar1}                     \\
     \frac{\mathrm{d}e}{\mathrm{d}t}      & = \frac{\sqrt{1-e^{2}}}{n_{o}a}\left[\sin{(f)}a_{R}+\left(\cos{f}+\frac{e+\cos{f}}{1+e\cos{f}}\right)a_{T}\right]           \\
     \frac{\mathrm{d}i}{\mathrm{d}t}      & = \frac{r \cos{u}}{n_{o}a^{2}\sqrt{1-e^{2}}}a_{N}                                                                           \\
     \frac{\mathrm{d}\Omega}{\mathrm{d}t} & = \frac{r \sin{u}}{n_{o}a^{2}\sqrt{1-e^{2}}\sin{i}} a_{N}                                                                   \\
     \frac{\mathrm{d}w}{\mathrm{d}t}      & = -\frac{\sqrt{1-e^{2}}}{n_{o}ae}\left[\cos{(f)}a_{R}-\left(\sin{f}+\frac{\sin{f}}{1+e\cos{f}}\right)a_{T}\right] \nonumber \\
                                          & -\frac{\mathrm{d}\Omega}{\mathrm{d}t}\cos{i}                                                                                \\
     \frac{\mathrm{d}f}{\mathrm{d}t}      & = \frac{h}{r^{2}} -\frac{\mathrm{d}w}{\mathrm{d}t}-\frac{\mathrm{d}\Omega}{\mathrm{d}t}\cos{i} \label{eq:GaussVar2}
\end{align}
where $n_{o}$ is the mean motion, $r$ is the orbital radius, $h$ is the orbital angular momentum, and $f$ is the true anomaly of the satellite being propagated.
The external accelerations $a_R, a_T$, and $a_N$ are expressed in the RTN frame.

\subsection{Light Curves}
\begin{figure}[tb]
     \centering
     \includegraphics[width=7cm]{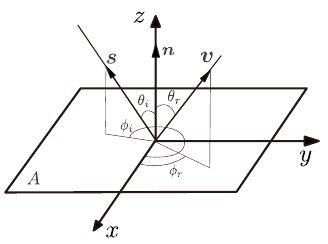}
     \caption{Light reflection geometry.}
     \label{fig:facet}
\end{figure}

Light curves are calculated by dividing the target object into many small facets and summing up the reflected light from each facet. The surface reflectance of each facet is modeled using the bidirectional reflectance distribution function (BRDF). Figure~\ref{fig:facet} illustrates the light reflection geometry on the target surface, where $\bm{s}$ and $\bm{v}$ represent the Sun directional vector and the reflection vector, respectively.
These vectors are defined using azimuth and polar angles $(\phi_{i}, \theta_{i})$ and $(\phi_{r}, \theta_{r})$.
The BRDF $f_{r}$ consists of a diffuse component $c_{d}$ and a specular component $c_{s}$ as follows:
\begin{align}
     f_{r} = c_{d} + c_{s}
\end{align}
In computer graphics, various BRDFs have been developed to represent physically based reflectances~\citep{overviewBRDF}.
Among these, this study employs the Ashikhmin--Shirley model, also known as the anisotropic Phong model~\citep{ashikhmin2000}. It is a physically based, energy-conserving BRDF that represents both diffuse and specular reflection with a small number of parameters and accounts for anisotropic surfaces, while remaining computationally efficient for facet-based light curve synthesis. These properties offer a favorable balance between fidelity and computational efficiency compared with the simpler Lambertian or Phong models, which motivates its wide use in space-object light curve modeling. The model is defined as
\begin{align}
     c_{d} & = \frac{28\rho_{d}}{23\pi}\left(1-F_{0}\right)\left[1-\left(1-\frac{\bm{n}^{T}\bm{s}}{2}\right)^{5}\right]\left[1-\left(1-\frac{\bm{n}^{T}\bm{v}}{2}\right)^{5} \right] \\
     c_{s} & = \frac{\sqrt{\left(n_{u}+1\right)\left(n_{v}+1\right)}}{8\pi}\frac{F}{\bm{v}^{T}\bm{h}\max{(\bm{n}^{T}\bm{s},\bm{n}^{T}\bm{v})}}(\bm{n}^{T}\bm{h})^{\alpha_{\rm AS}}
\end{align}
where $\rho_{d}$ is the diffuse reflectance and
\begin{align}
     \alpha_{\rm AS}  = {n_{u}\cos^{2}{\phi_{h}}+n_{v}\sin^{2}{\phi_{h}}}
     = \frac{n_{u}(\bm{h}^{T}\bm{u}_{u})^{2} + n_{v}(\bm{h}^{T}\bm{u}_{v})^{2}}{1-(\bm{h}^{T}\bm{n})^{2}}
\end{align}
The vector $\bm{h}$ is the bisector between the Sun directional vector $\bm{s}$ and the reflection vector $\bm{v}$, defined as
\begin{align}
     \bm{h}  = \frac{\bm{s}+\bm{v}}{\|\bm{s}+\bm{v}\|}
     =\begin{bmatrix}
           \cos{\phi_{h}}\sin{\theta_{h}} \\
           \sin{\phi_{h}}\sin{\theta_{h}} \\
           \cos{\theta_{h}}
      \end{bmatrix}
\end{align}
The unit directional vectors $\bm{u}_{u}$ and $\bm{u}_{v}$ define the anisotropy direction, and the magnitudes of anisotropy are described by $n_{u}$ and $n_{v}$.
Fresnel reflectance $F$ is expressed as
\begin{align}
     F=F_{0}+\left(1-F_{0}\right)(1-\bm{v}^{T}\bm{h})^{5}
\end{align}
where $F_{0}$ is the Fresnel reflectance at normal incidence.

The light intensity on the $i$th facet $f_{{\rm obs},i}$ is written as
\begin{align}
     f_{{\rm obs},i } = \frac{C_{\rm sun}f_{r,i} (\bm{n}_{i}^{T}\bm{s}) (\bm{n}_{i}^{T}\bm{v}) A_{i}}{r^{2}_{\rm obs}} \label{eq:fobs}
\end{align}
where the subscript $i$ indicates the $i$th facet, $C_{\rm sun}$ is the light intensity of the Sun, $A_{i}$ is the facet area, and $r_{\rm obs}$ is the distance between the target object and the observer.
Note that $f_{{\rm obs},i}=0$ when $\bm{n}_{i}^{T}\bm{s}$ and/or $\bm{n}_{i}^{T}\bm{v}$ are negative, indicating that the object's surface is in eclipse or not oriented toward the observer, respectively.

Earth albedo~\cite{wettererRefiningSpaceObject2014} is also incorporated into the light curve computation.
Earth is divided into grid points that have equal area, and the irradiance at the target from the $j$-th grid point $\mathrm{d}E_{j}$ is calculated as
\begin{align}
     \mathrm{d}E_{j} = \frac{C_{\rm sun} A_j}{\pi r_{j}^{2}} a_{\oplus} \cos{\theta_{n_j, t}}\cos{\theta_{n_j, s}}
\end{align}
where $A_j=\frac{4\pi R_\oplus^2}{N_{\oplus}}$ is the area of the $j$-th grid point, $R_\oplus$ is the Earth radius, $N_{\oplus}$ is the number of grid points, $a_\oplus$ is the albedo of the Earth, $r_j$ is the distance between the target and the $j$-th grid point, $\theta_{n_j, t}$ is the angle between the target direction and the normal vector of the $j$-th grid point, and $\theta_{n_j, s}$ is the angle between the Sun direction and the normal vector of the $j$-th grid point. This paper assumes a constant Earth albedo of $0.3$, which is a representative globally averaged value for Earth~\cite{montenbruckGill2000}.
The light intensity on the $i$-th facet reflected from Earth albedo, $f_{{\rm obs},i,\oplus}$, is calculated by summing the contribution from all visible Earth grid points:
\begin{align}
     f_{{\rm obs},i,\oplus} = \sum_{j=1}^{N_{\oplus, {\rm visible}}} \frac{\mathrm{d}E_{j} f_{r,i,j} (\bm{n}_{i}^{T}\bm{{s}}_{j}) (\bm{n}_{i}^{T}\bm{v}) A_{i}}{r_{\rm obs}^{2}}
\end{align}
where $N_{\oplus, {\rm visible}}$ is the number of visible grid points from the target, $\bm{{s}}_{j}$ is the unit directional vector from the $i$-th facet of the target to the $j$-th grid point of Earth, and $f_{r,i,j}$ is the BRDF for the incoming direction $\bm{{s}}_{j}$.

The sum of $f_{{\rm obs}, i}$ (due to direct solar radiation) and $f_{{\rm obs},i,\oplus}$ (due to Earth albedo) for all the facets provides the total observed flux.
The apparent magnitude of the light curves is written as
\begin{equation}
     m_{\text{app}} = m_{\text{sun}} - 2.5 \log_{10}\frac{{ \sum\limits_{i=1}^N}\textstyle \left( f_{\text{obs},i} + f_{{\rm obs},i,\oplus} \right)}{C_{\mathrm{sun}}} \label{eq:mapp}
\end{equation}
where $m_{\rm sun}=-26.7$ is the apparent magnitude of the Sun and $N$ is the number of facets.
Non-convex objects exhibit self-shadowing depending on illumination conditions. In this study, self-shadowing is accounted for using a ray-tracing method.

\subsection{Attitude Representation}\label{subsec:attitude}
Light curves are highly nonlinear functions of the target attitude.
In the numerical simulations, the target attitude used to calculate the light curves is represented by quaternions and propagated with the standard quaternion kinematics and Euler's rigid-body equations~\citep{crassidisJunkins2012}.
% The error quaternion $\bm{q}_{e}$ between the true quaternion $\bm{q}_{\rm true}$ and the estimate $\hat{\bm{q}}$ is defined as
% \begin{align}
% \bm{q}_{e} = \bm{q}_{\rm true}\otimes \hat{\bm{q}}^{\dagger}
% \end{align}
% where $\hat{\bm{q}}^{\dagger}$ is the conjugate of $\hat{\bm{q}}$ defined as $\bm{q}^{\dagger} =[-\bar{\bm{q}}^{T}, q_{4}]^{T}$.
% Although quaternion has no singularity to represent attitude, quaternion has norm constraint as
% \begin{align}
% \|\bm{q}\|^{2} = \bm{q}^{T}\bm{q} = 1
% \end{align}
% This constraint is violated in weighted averaging in the UKF.
% Thus the attitude representation is transformed to the generalized Rodrigues parameter (GRP) $\bm{p}$ as
% \begin{align}
% \bm{p} = f_{\rm GRP}\frac{\bar{\bm{q}}}{a_{\rm GRP}+q_{4}}
% \end{align}
% where $a_{\rm GRP}\in [0,1]$ and $f_{\rm GRP}$ is a scale factor.
% The detailed UKF procedure using the GRP can be found in~~\cite{crassidis2003a}.

For simplicity, it is assumed that the body-fixed frame coincides with the principal axes of inertia and that no external torque acts on the satellite.
Although the use of light curve measurements enables the simultaneous estimation of attitude and ROE, this paper assumes that the target attitude is known because the purpose of this paper is to investigate the effectiveness of light curves in improving the observability.
This assumption isolates the observability contribution of the light curve to the relative state, so that the gain is attributable to the recovered range information rather than to simultaneous attitude estimation, and it renders the brightness modulation predictable so that the model-independent range term of Eq.~\eqref{eq:mappsep} can be cleanly extracted.
It is not required for the filter to operate, as the same framework supports the joint estimation of attitude and ROE, which is left for future work.
The assumption is reasonable for a meaningful class of targets whose attitude is externally available, such as three-axis or nadir-stabilized satellites, objects already characterized on an earlier pass, or targets close enough to be resolved by imaging. The targets demonstrated here nonetheless tumble, and the method is tested against full attitude-driven brightness variation rather than a conveniently stabilized signal.

\section{Method}
\subsection{Observability Analysis}\label{subsec:obs}
This paper considers bearing angles $(\alpha, \beta)$ and the apparent magnitude of the light curve $m_{\rm app}$ for relative orbit estimation, which are summarized as
\begin{align}
     \bm{y} & = \begin{bmatrix}
                     \alpha \\
                     \beta  \\
                     m_{\rm app}
                \end{bmatrix} + \bm{\nu}  \\
            & = \bm{h}(\bm{x}) + \bm{\nu}
\end{align}
where $\bm{x}= \delta\bi{\oe}$, $\bm{h}(\bm{x})$ is the measurement function defined with Eqs.~\eqref{eq:azi},~\eqref{eq:ele}, and~\eqref{eq:mapp}, and $\bm{\nu}=[\nu_{\rm LOS},\nu_{\rm LOS}, \nu_{\rm LC}]^{T}$ is the measurement noise vector.
The measurement noises $\nu_{\rm LOS}$ and $\nu_{\rm LC}$ are assumed to follow the normal distribution with zero mean and standard deviations $\sigma_{\rm LOS}$ and $\sigma_{\rm LC}$, respectively.

The observability analysis is evaluated with the Fisher information matrix $\mathcal{I}(\bm{x})$ defined as
\begin{align}
     \mathcal{I}(\bm{x}) & = \mathbb{E}\left\{\left[\frac{\partial}{\partial \bm{x}} \ln{p(\bm{y}|\bm{x})}\right]\left[\frac{\partial }{\partial \bm{x}} \ln{p(\bm{y}|\bm{x})}\right]^{T}\right\}
\end{align}
where $\mathbb{E}\{\cdot\}$ denotes the expectation. The Fisher information matrix is reduced to the following form due to the additive measurement noise
\begin{align}
     \mathcal{I}(\bm{x}) & = \mathbb{E}\left[{H}^{T}{R}^{-1}{H}\right]
\end{align}
where $H$ is the Jacobian matrix of the measurement function $\bm{h}(\bm{x})$ and $R$ is the measurement noise covariance matrix.
The Jacobian matrix is decomposed as
\begin{align}
     H = \begin{bmatrix}
              H_{\alpha,\beta} \\
              H_{m_{\rm app}}
         \end{bmatrix}
\end{align}
where $H_{\alpha,\beta}$ is the Jacobian matrix with respect to the bearing angles and $H_{m_{\rm app}}$ is the Jacobian matrix with respect to the light curve.
The Jacobian $H_{\alpha,\beta}$ is calculated~\cite{Sullivan2017,sumi2025iac} as
\begin{align}
     H_{\alpha,\beta} & = \left.\frac{\partial \bm{h}}{\partial \delta \bi{\oe}} \right|_{\delta \bi{\oe}}
     = \left.\frac{\partial \bm{h}}{\partial \delta\bm{r}^{\mathcal{B}}_{d}} \frac{\partial  \delta\bm{r}_{d}^{\mathcal{B}}}{\partial \delta \bm{r}_{d}^{\mathcal{I}}} \frac{\partial \delta \bm{r}_d^{\mathcal{I}}}{\partial \delta \bm{r}_{d}^{\mathcal{O}}} \frac{\partial \delta \bm{r}_{d}^{\mathcal{O}}}{\partial \delta \bi{\oe}}\right|_{\delta \bi{\oe}} \\
                      & =\left.\frac{\partial \bm{h}}{\partial \delta\bm{r}^{\mathcal{B}}_{d}} R_{b / i} R_{i / o} \frac{\partial \delta \bm{r}_{d}^{\mathcal{O}}}{\partial \delta \bi{\oe}}\right|_{\delta \bi{\oe}}
\end{align}
where
\begin{align}
     \frac{\partial \bm{h}}{\partial \delta\bm{r}^{\mathcal{B}}_{d}} = \frac{1}{\|\delta \bm{r}^{\mathcal{B}}_{d}\|} \begin{bmatrix}
                                                                                                                          -\frac{\sin{\alpha}}{\cos{\beta}} & \frac{\cos{\alpha}}{\cos{\beta}} & 0           \\
                                                                                                                          - \cos{\alpha}\sin{\beta}         & - \sin{\alpha}\sin{\beta}        & \cos{\beta}
                                                                                                                     \end{bmatrix}
\end{align}
and $R_{b / i}$ and $R_{i / o}$ are the rotation matrices from the inertial frame to the body-fixed frame and from the orbital frame to the inertial frame, respectively.
The mapping from ROE to the relative position in the RTN frame is given by
\begin{align}
     \frac{\partial \delta \bm{r}_{d}^{\mathcal{O}}}{\partial \delta \bi{\oe}} = a\begin{bmatrix}
                                                                                       1 & 0 & -\cos{u} & -\sin{u}  & 0       & 0        \\
                                                                                       0 & 1 & 2\sin{u} & -2\cos{u} & 0       & 0        \\
                                                                                       0 & 0 & 0        & 0         & \sin{u} & -\cos{u}
                                                                                  \end{bmatrix}
\end{align}
The Jacobian with respect to the light curve $m_{\rm app}$ is written as
\begin{align}
     H_{m_{\rm app}} & = \frac{\partial m_{\rm app}}{\partial \delta\bm{r}^\mathcal{B}_{d}} \frac{\partial \delta\bm{r}^\mathcal{B}_{d}}{\partial \delta \bm{r}^\mathcal{O}_{d}} \frac{\partial \delta \bm{r}^\mathcal{O}_{d}}{\partial \delta \bi{\oe}} \\
                     & = \frac{\partial m_{\rm app}}{\partial \delta\bm{r}^\mathcal{B}_{d}} R_{b/o} \frac{\partial \delta \bm{r}^\mathcal{O}_{d}}{\partial \delta \bi{\oe}}
\end{align}
where
\begin{align}
     \frac{\partial m_{\rm app}}{\partial \delta\bm{r}^\mathcal{B}_{d}} = \frac{5}{\|\delta \bm{r}^\mathcal{B}_{d}\|^2\ln{10}} \delta\bm{r}^\mathcal{B}_{d}
\end{align}
This gradient reveals that the observability contributed by the light curve is independent of the detailed photometric model. Writing the total observed flux as $S=\sum_{i}(f_{{\rm obs},i}+f_{{\rm obs},i,\oplus})=G/r_{\rm obs}^{2}$, where the level $G$ collects the reflection of all facets, which depends on the attitude, shape, and optical properties, and $r_{\rm obs}=\|\delta\bm{r}^\mathcal{B}_{d}\|$ is the observer--target range, the apparent magnitude in Eq.~\eqref{eq:mapp} separates as
\begin{align}
     m_{\rm app} = \underbrace{\left[m_{\rm sun}-2.5\log_{10}\frac{G}{C_{\rm sun}}\right]}_{\text{absolute level}} + \underbrace{5\log_{10} r_{\rm obs}}_{\text{range term}} \label{eq:mappsep}
\end{align}
The level $G$ enters additively and cancels under the derivative of Eq.~\eqref{eq:mappsep} with respect to $r_{\rm obs}$, leaving the range gradient $\partial m_{\rm app}/\partial r_{\rm obs}=5/(r_{\rm obs}\ln 10)$, so the gradient that feeds $H_{m_{\rm app}}$ contains no attitude, shape, or optical parameter. Consequently, the information through which the light curve restores the observability of the relative range is model-independent, which is why approximate optical properties do not corrupt the recovered relative state, and the diffuse reflectance $\rho_{d}$ can be estimated online rather than assumed known.

Because the target attitude is treated as a known time-varying input, it does not appear in the estimated state. Accordingly, the observability Jacobian $H_{m_{\rm app}}$ contains only the partial derivatives with respect to the relative state. The attitude-induced brightness variation is thus part of the deterministic measurement model, neither neglected nor treated as measurement noise.

The inverse of the Fisher information matrix provides a lower bound on the covariance of any unbiased estimator. The diagonal elements of this inverse matrix are used to represent the minimum achievable variance, and the following index is used to evaluate observability in Section~\ref{sec:num}.
\begin{align}
     \mathcal{I} = \sqrt{\mathcal{I}_{k,c}^{-1}(\bm{x})} \label{eq:observability}
\end{align}
where $\mathcal{I}_{k,c}(\bm{x})$ is the cumulative Fisher information matrix defined as
\begin{align}
     \mathcal{I}_{k,c}(\bm{x}) = \sum_{i=0}^{k} \mathcal{I}_{i}(\bm{x}) \label{eq:fisher}
\end{align}
% The subscript $k$ in Eq.~\eqref{eq:fisher} indicates the $k$-th time instant.
A smaller value of the index indicates a smaller minimum variance, thus indicating higher observability.
It is noted that the Fisher information matrix is formed for the relative state $\delta\bi{\oe}$ only. The photometric parameters, including the diffuse reflectance $\rho_{d}$, are treated as known in the Jacobian. The resulting index is therefore a known-optics observability bound, whereas the realized performance when $\rho_{d}$ is unknown is obtained from the Monte Carlo estimation with online $\rho_{d}$ augmentation.

\subsection{Adaptive Unscented Kalman Filter}\label{subsec:ukf}
This paper uses the adaptive unscented Kalman filter to estimate the relative orbit. The adaptive process noise can accommodate the uncertainty of the external disturbances. In addition, the state variable is augmented with the diffuse reflectance $\rho_d$ to enhance the robustness of the estimation against the uncertainty of the reflectance. The choice of which reflectance to estimate follows from how each enters the light curve. The diffuse reflectance sets the persistent brightness level at essentially every epoch, so an incorrectly assumed value displaces the predicted magnitude at all times. Through the range term of Eq.~\eqref{eq:mappsep}, it maps into a proportional bias of the estimated relative orbit. This effect is quantified in Section~\ref{sec:num}. The specular reflectance, in contrast, contributes appreciably only in limited geometrical conditions, known as glint~\cite{matsushitaConceptualStudyImproved2024}, and is therefore held at its nominal value rather than estimated. Glint epochs are not discarded, they are simply rare, and remain part of the deterministic measurement model. Although the apparent magnitude adds only a single scalar per measurement epoch, the augmented parameter $\rho_d$ is constant, so the entire light curve arc accumulates information about it alongside the relative state. It is noted that the primary contribution of this paper is to investigate the effectiveness of light curves in improving the observability, rather than the development of novel estimation algorithms.
The following estimation procedure employs established techniques and is not a new contribution of this paper.
\begin{enumerate}
     \item Let $\bm{x}=[\delta \bi{\oe}^T,\rho_d]^T$ be the state variable, $\hat{\bm{x}}_{t_{0}}$ be the initial estimate, and $P_{t_{0}}$ be the initial covariance.
     \item For $n$ state variables ($n=7$ in this paper), $2n+1$ sigma points $\bm{\chi}$ at time $t_{k}$ are calculated as
           \begin{align}
                \bm{\chi}_{0,t_{k}}   & = \hat{\bm{x}}_{t_{k}}                                                           \\
                \bm{\chi}_{i,t_{k}}   & = \hat{\bm{x}}_{t_{k}} +\sqrt{n+\lambda}(\sqrt{P_{t_{k}}})_{i},~(i=1,2,\dots,n)  \\
                \bm{\chi}_{n+i,t_{k}} & = \hat{\bm{x}}_{t_{k}} - \sqrt{n+\lambda}(\sqrt{P_{t_{k}}})_{i},~(i=1,2,\dots,n)
           \end{align}
           where $\left(\sqrt{P}_{t_{k}}\right)_{i}$ is the $i$th column vector of $\sqrt{P}_{t_{k}}$.
           The parameter $\lambda = \alpha_{\rm UKF}^{2}(n+\kappa) - n$ is a composite scaling parameter, where $\alpha_{\rm UKF}$ and $\kappa$ are tuning parameters that determine the spread of the sigma points around the mean state~\citep{crassidisJunkins2012}. Here $\kappa$ is set to zero rather than the common heuristic $\kappa=3-n$, which would be negative for the present state dimension and could compromise the positive semidefiniteness of the covariance, and $\alpha_{\rm UKF}$ is set to a small positive value to keep the sigma points close to the mean (see Table~\ref{tab:ukfPara}).
     \item Each sigma point is propagated to the time $t_{k+1}$ as
           \begin{align}
                \bm{\chi}^{-}_{i,t_{k+1}} = \bm{f}(\bm{\chi}_{i,t_{k}}),~(i=0,1,\dots,2n)
           \end{align}
           % Since the state variable is defined with the error GRP, it is transformed to quaternions for state propagation. 
           The state propagation is calculated with the GVE in Eqs.~\eqref{eq:GaussVar1}--\eqref{eq:GaussVar2}.
     \item The a priori state estimate $\hat{\bm{x}}^{-}_{t_{k+1}}$ and a priori covariance $P^{-}_{t_{k+1}}$ are calculated as
           \begin{align}
                \hat{\bm{x}}^{-}_{t_{k+1}} & = \sum_{i=0}^{2n}w_{i}^{m}\bm{\chi}^{-}_{i,t_{k+1}}                                                                                                                 \\
                P^{-}_{t_{k+1}}            & = \sum_{i=0}^{2n}w_{i}^{c}(\bm{\chi}^{-}_{i,t_{k+1}} - \hat{\bm{x}}^{-}_{t_{k+1}})(\bm{\chi}^{-}_{i,t_{k+1}} - \hat{\bm{x}}^{-}_{t_{k+1}})^{T} + Q \label{eq:Pstep}
           \end{align}
           where $Q$ is the process noise and the weights are obtained as
           \begin{align}
                w^{m}_{0} & = \frac{\lambda}{n+\lambda}                                                             \\
                w^{m}_{i} & = \frac{1}{2(n+\lambda)}, ~(i=1,2,\dots,2n)                                             \\
                w^{c}_{0} & = \frac{\lambda}{n+\lambda} + 1 - \alpha_{\rm UKF}^{2} + \beta_{\rm UKF} \label{eq:wc0} \\
                w^{c}_{i} & = \frac{1}{2(n+\lambda)}, ~(i=1,2,\dots,2n)
           \end{align}
           In Eq.~\eqref{eq:wc0}, $\beta_{\rm UKF}$ is a tuning parameter that incorporates prior knowledge of the distribution of the state; $\beta_{\rm UKF}=2$ is optimal for a Gaussian distribution and is adopted here (see Table~\ref{tab:ukfPara}).
     \item The predicted measurements are calculated from each sigma point using Eqs.~\eqref{eq:azi},~\eqref{eq:ele}, and~\eqref{eq:mapp} as
           \begin{align}
                \mathcal{\bm{Y}}^{-}_{i,t_{k+1}} = \bm{h}(\bm{\chi}^{-}_{i,t_{k+1}}),~(i=0,1,\dots,2n)
           \end{align}
           The mean predicted measurement $\hat{\bm{y}}^{-}$, covariance matrix $P_{yy,t_{k+1}}^{-}$, and correlation matrix $P^{-}_{xy,t_{k+1}}$ are obtained as
           \begin{align}
                \hat{\bm{y}}^{-}_{t_{k+1}} & = \sum_{i=0}^{2n}w_{i}^{m}\mathcal{\bm{Y}}^{-}_{i,t_{k+1}}                                                                                                   \\
                P^{-}_{yy, t_{k+1}}        & = \sum_{i=0}^{2n}w_{i}^{c}(\mathcal{\bm{Y}}^{-}_{i,t_{k+1}} - \hat{\bm{y}}^{-}_{t_{k+1}})(\mathcal{\bm{Y}}^{-}_{i,t_{k+1}} - \hat{\bm{y}}^{-}_{t_{k+1}})^{T} + R \\
                P^{-}_{xy, t_{k+1}}        & = \sum_{i=0}^{2n}w_{i}^{c}({\bm{\chi}}^{-}_{i,t_{k+1}} - \hat{\bm{x}}^{-}_{t_{k+1}})(\mathcal{\bm{Y}}^{-}_{i,t_{k+1}} - \hat{\bm{y}}^{-}_{t_{k+1}})^{T}
           \end{align}
           where $R$ is the observation noise matrix.
     \item The filtered state and covariance matrix are obtained as
           \begin{align}
                \hat{\bm{x}}_{t_{k+1}} & = \hat{\bm{x}}^{-}_{t_{k+1}} + K(\bm{y}_{t_{k+1}} - \hat{\bm{y}}^{-}_{t_{k+1}}) \label{eq:measUpdate} \\
                P_{t_{k+1}}            & = P^{-}_{t_{k+1}} - K P^{-}_{yy,t_{k+1}}K^{T}
           \end{align}
           where $K$ is the Kalman gain defined as
           \begin{align}
                K = P^{-}_{xy,t_{k+1}} (P^{-}_{yy,t_{k+1}})^{-1}
           \end{align}
\end{enumerate}
Two measurement-editing rules are applied in the update step. First, epochs with no optical observation, where the target is in the Earth's shadow or presents no sunlit facet toward the observer, are propagated without a measurement update. Second, the apparent magnitude in Eq.~\eqref{eq:mapp} is singular as the received flux approaches zero: $m_{\rm app}\to\infty$ at the boundary of the visibility condition, where the measurement sensitivity diverges. Epochs at which the observed or sigma point predicted magnitude is fainter than $m_{\rm app}=20$, far below any practical photometric detection limit, or at which the predicted magnitude spreads by more than one magnitude across the sigma points, indicating that the measurement model is locally discontinuous, are therefore processed as bearing-only updates with the light curve component discarded. This editing reflects the finite sensitivity of a real photometric sensor and protects the filter from the logarithmic singularity of the magnitude scale near zero flux.

Although the target shape and attitude are assumed known, so that the measurement model is well specified, setting an appropriate constant value for the process noise $Q$ in the fourth step is still difficult because $Q$ must compensate for the mismodeling of the relative orbital dynamics rather than for the measurement. The filter propagates the state with the Gauss variational equations, which do not reproduce all the perturbations acting on the true relative motion, such as the higher-order terms of the Earth's gravity field, solar radiation pressure whose magnitude depends on the optical property being estimated, and the third-body attractions of the Sun and Moon. The statistics of the resulting dynamics errors are orbit dependent and not known a priori. This paper therefore uses an adaptive method~\citep{Mohamed}, which is also known as covariance matching~\citep{Meng}.
The innovation is defined as
\begin{align}
     \Delta \bm{y}_{k+1} = \bm{y}_{t_{k+1}} -\hat{\bm{y}}^{-}_{t_{k+1}}
\end{align}
Using the measurement update in Eq.~\eqref{eq:measUpdate}, the residual is rewritten as
\begin{align}
     \Delta \bm{x} = K(\bm{y}_{t_{k+1}} - \hat{\bm{y}}^{-}_{t_{k+1}})
\end{align}
The process noise covariance is obtained as
\begin{align}
     Q & = \mathbb{E}[\Delta\bm{x}\Delta\bm{x}^{T}]                                                                                               \\
       & = K \mathbb{E}[(\bm{y}_{t_{k+1}} - \hat{\bm{y}}^{-}_{t_{k+1}})(\bm{y}_{t_{k+1}} - \hat{\bm{y}}^{-}_{t_{k+1}})^{T}] K^{T} \label{eq:Qest}
\end{align}
The innovation term in Eq.~\eqref{eq:Qest} is approximated using the last $N_{\rm innov}$ innovations as
\begin{align}
     \mathbb{E}[(\bm{y}_{t_{k+1}} - \hat{\bm{y}}^{-}_{t_{k+1}})(\bm{y}_{t_{k+1}} - \hat{\bm{y}}^{-}_{t_{k+1}})^{T}] \approx \frac{1}{N_{\rm innov}}\sum^{N_{\rm innov}}_{i=k-N_{\rm innov}+1}(\bm{y}_{t_{i+1}} - \hat{\bm{y}}^{-}_{t_{i+1}})(\bm{y}_{t_{i+1}} - \hat{\bm{y}}^{-}_{t_{i+1}})^{T} \label{eq:Ninnovation}
\end{align}
Thus the process noise covariance with Eqs.~\eqref{eq:Qest} and~\eqref{eq:Ninnovation} is used in Eq.~\eqref{eq:Pstep}.

\section{Numerical Examples}\label{sec:num}
\begin{figure}[tb]
     \centering
     \includegraphics{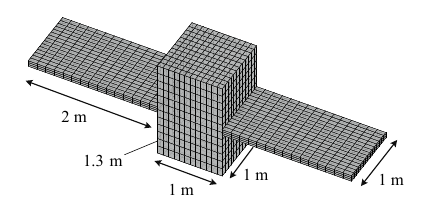}
     \caption{Size of a box-wing object.}
     \label{fig:boxWing}
\end{figure}

Monte Carlo simulation with $100$ samples is performed to verify the effectiveness of the light curve measurement.
Two target objects are considered: a flat plate object and a box-wing object.
The size of the flat plate is $1~{\rm m}\times 1~{\rm m}$ and that of the box-wing object is illustrated in Fig.~\ref{fig:boxWing}.
The moments of inertia are $J = {\rm diag}(0.8, 0.8, 1.6)~{\rm kg\,m^2}$ and $J= {\rm diag}(35, 70, 80)~{\rm kg\,m^2}$ for the flat plate and the box-wing object, respectively. It is noted that the box-wing object has a non-convex shape, and self-shadowing is considered in the numerical simulation.
The optical parameters are $(n_{u},n_{v}) = (800, 800)$, $F_{0} = 0.5$, and $\rho_{d}=0.5$ for both objects.
These optical parameters are held constant across all Monte Carlo runs. The reflectances $F_{0}$ and $\rho_{d}$ are set to the middle of their physical range $[0,1]$, and the exponents $(n_{u},n_{v})$ correspond to a moderately concentrated isotropic specular lobe, so that the simulated light curves contain both a persistent diffuse component and attitude-dependent specular glints.
The conclusions are not tied to these particular values because the range information supplied by the light curve in Eq.~\eqref{eq:mappsep} is independent of the optical parameters, which set only the absolute brightness level. Robustness to uncertainty in the brightness level is nevertheless verified by treating the diffuse reflectance $\rho_{d}$ as unknown and estimating it jointly with the relative state.
The sensitivity of the estimation performance to the remaining scenario choices, namely the initial relative orbit, the illumination conditions, and the fidelity of the target shape model, is examined separately in Section~\ref{subsec:sens}.
The absolute orbit of the chief and the relative orbit of the deputy are summarized in Table~\ref{tab:inicon}.
This orbital condition is challenging because the target satellite enters the Earth's shadow, where no optical observation is available.
The absolute orbit of the chief as a true value is calculated with GVE including the 8th order and degree of Earth gravity model, solar radiation pressure, and the third-body gravity perturbations by Moon and Sun. The angular rate of the target satellite is set to $\bm{\omega}=[0.05,~0.05,~0.01]^{T}$ rad/s. The estimation parameters in Table~\ref{tab:ukfPara} are used for all test cases.

Each Monte Carlo run randomizes three elements of the problem. First, the initial attitude of the target is sampled from the uniform distribution; since the apparent magnitude depends on the attitude history, each run processes a different light curve realization. Second, an independent realization of the measurement noise in Table~\ref{tab:ukfPara} is drawn for the bearing angles and the apparent magnitude. Third, the initial state estimate is generated by scaling the entire true initial ROE vector by a single random factor with a maximum error of 30\%. A common scaling of all ROE components changes the size of the relative orbit while preserving the line-of-sight directions, so this initialization places the initial error along the range-ambiguous direction that bearing measurements cannot correct, directly testing whether the light curve resolves the range ambiguity.

A run is counted as converged when the estimate remains finite and bounded at the final time. The effectiveness of the light curve measurement is quantified by three metrics over the 100 runs: the convergence rate, the root-mean-square error and standard deviation of the converged runs, and the convergence time expressed in orbital periods. Because the angles-only and the light curve augmented filters are evaluated under the same randomization scheme, a consistent improvement in these metrics is attributable to the added light curve measurement rather than to a particular attitude history, noise realization, or initial estimate.

\begin{table}[tb]
     \begin{center}
          \caption{UKF parameters and noise values} \label{tab:ukfPara}
          \begin{tabular}{lccc}
               \hline \hline
               noise          & $\sigma_{\rm LOS}~{\rm arcsec}$ & $\sigma_{\rm LC}$ &          \\ \hline
                              & $30$                            & $10^{-1}$         &          \\ \hline
               UKF parameters & $\alpha_{\rm UKF}$              & $\beta_{\rm UKF}$ & $\kappa$ \\ \hline
                              & $10^{-4}$                       & $2$               & $0$      \\
               \hline \hline
          \end{tabular}
     \end{center}
\end{table}

\begin{table*}[tb]
     \begin{center}
          \caption{Absolute orbital elements of chief and relative orbital elements of deputy} \label{tab:inicon}
          \begin{tabular}{lcccccc}
               \hline \hline
               Chief  & Altitude~km    & $e_{x}$               & $e_{y}$                & $i~{\rm deg}$           & $\Omega~{\rm deg}$ & $u_{0}$~deg \\
               \hline
                      & $850.0$        & $-5.00\times 10^{-4}$ & $8.66\times 10^{-4}$   & $98.700$                & $60.000$           & $0.000$     \\

               % GEO    & 35786.0        & $-5.00\times 10^{-4}$ & $8.66\times 10^{-4}$   & 1.000                   & 60.000             & 300.000     \\
               \hline
               Deputy & $a\delta a$ km & $a\delta\lambda$ km   & $a\|\delta\bm{e}\|$ km & $a\|\delta \bm{i}\|$ km &                    &             \\
               \hline
                      & $0.000$        & $-30.000$             & $0.500$                & $0.500$                 &                    &             \\
               \hline \hline
          \end{tabular}
     \end{center}
\end{table*}

\subsection{Flat Plate Target}
\begin{figure}[tb]
     \centering
     \includegraphics[width=11cm]{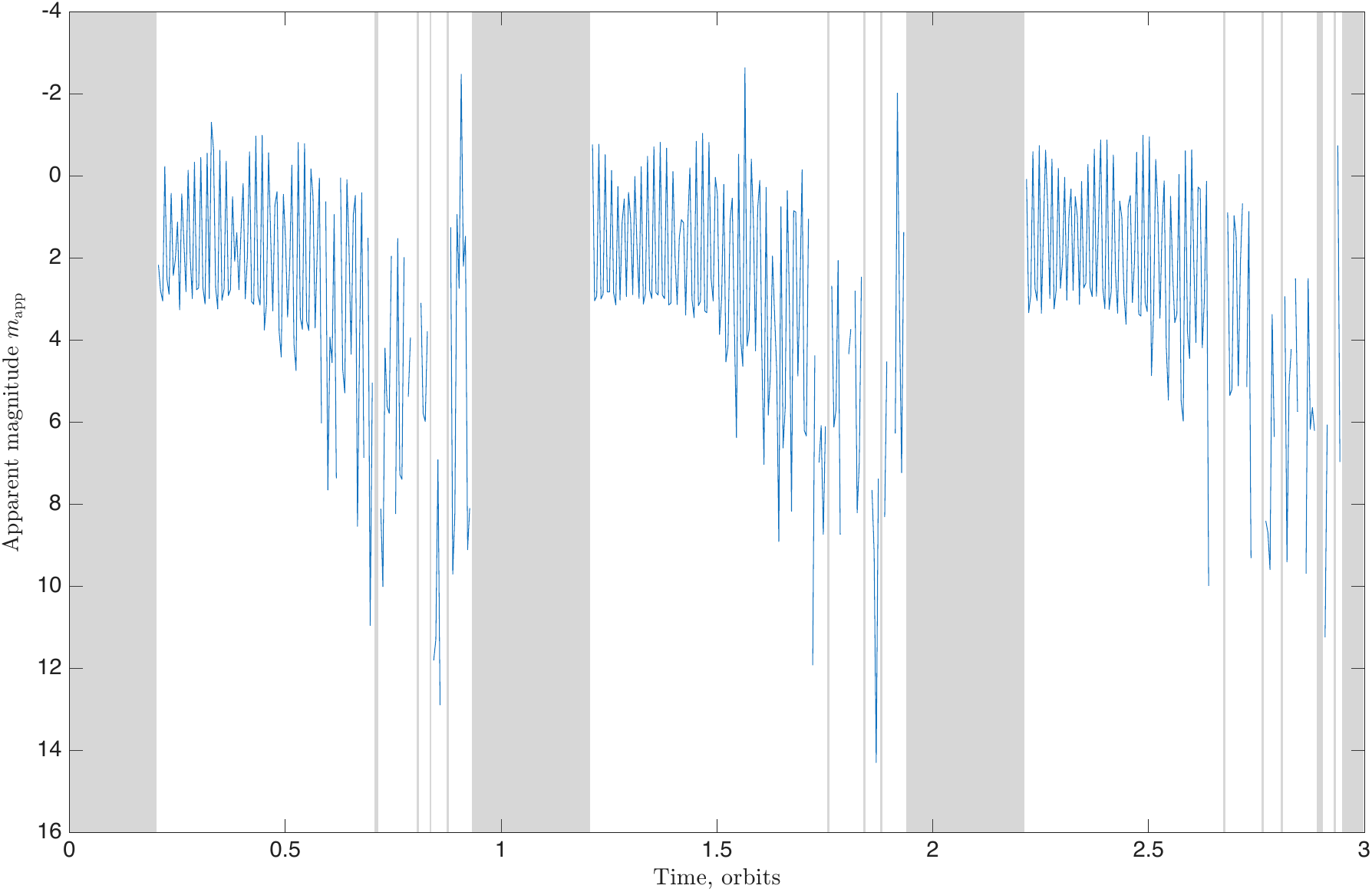}
     \caption{Light curves of flat plate target in LEO.}
     \label{fig:lcFlatPlateLEO}
\end{figure}

\begin{figure}[tb]
     \centering
     \includegraphics[width=12cm]{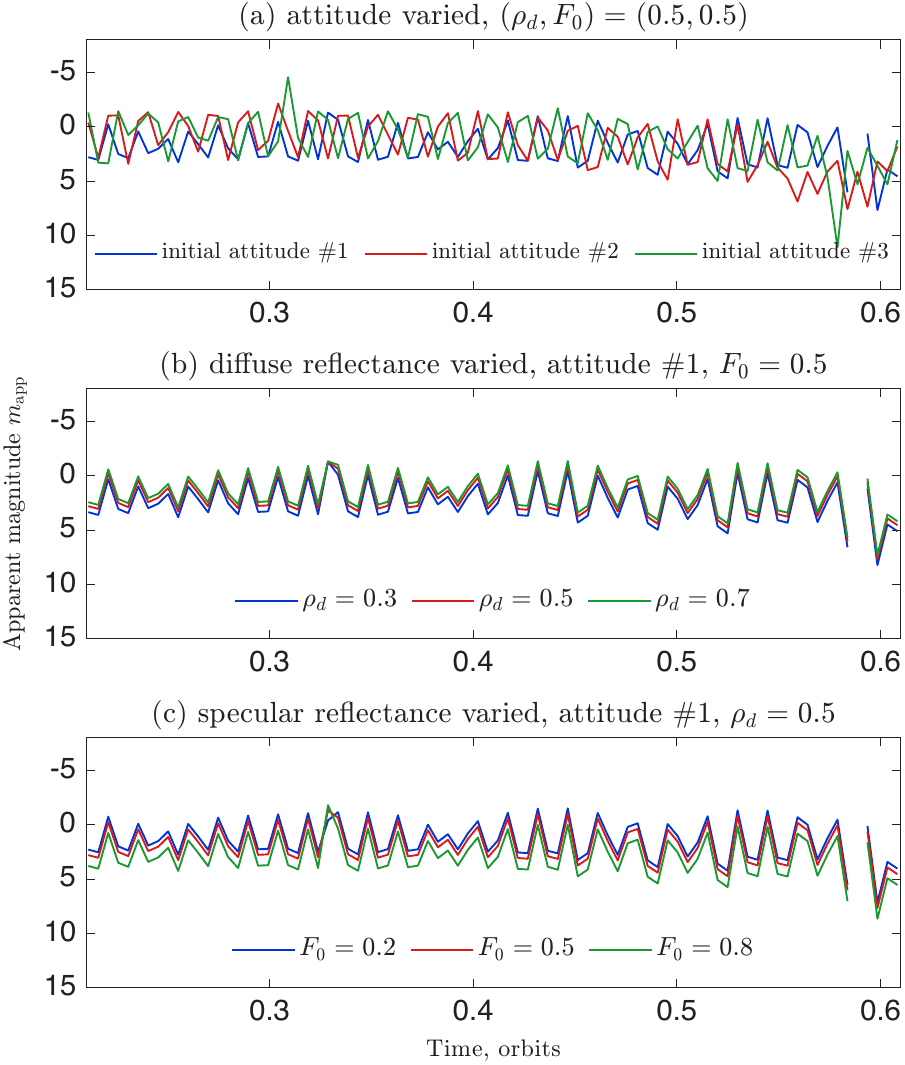}
     \caption{Sensitivity of the light curve to the target attitude and optical properties. The attitude reshapes the fast modulation, whereas the optical properties shift the brightness level. The range term of Eq.~\eqref{eq:mappsep} is common to all curves.}
     \label{fig:lcSensitivity}
\end{figure}

\begin{figure}[tb]
     \centering
     \includegraphics[width=12cm]{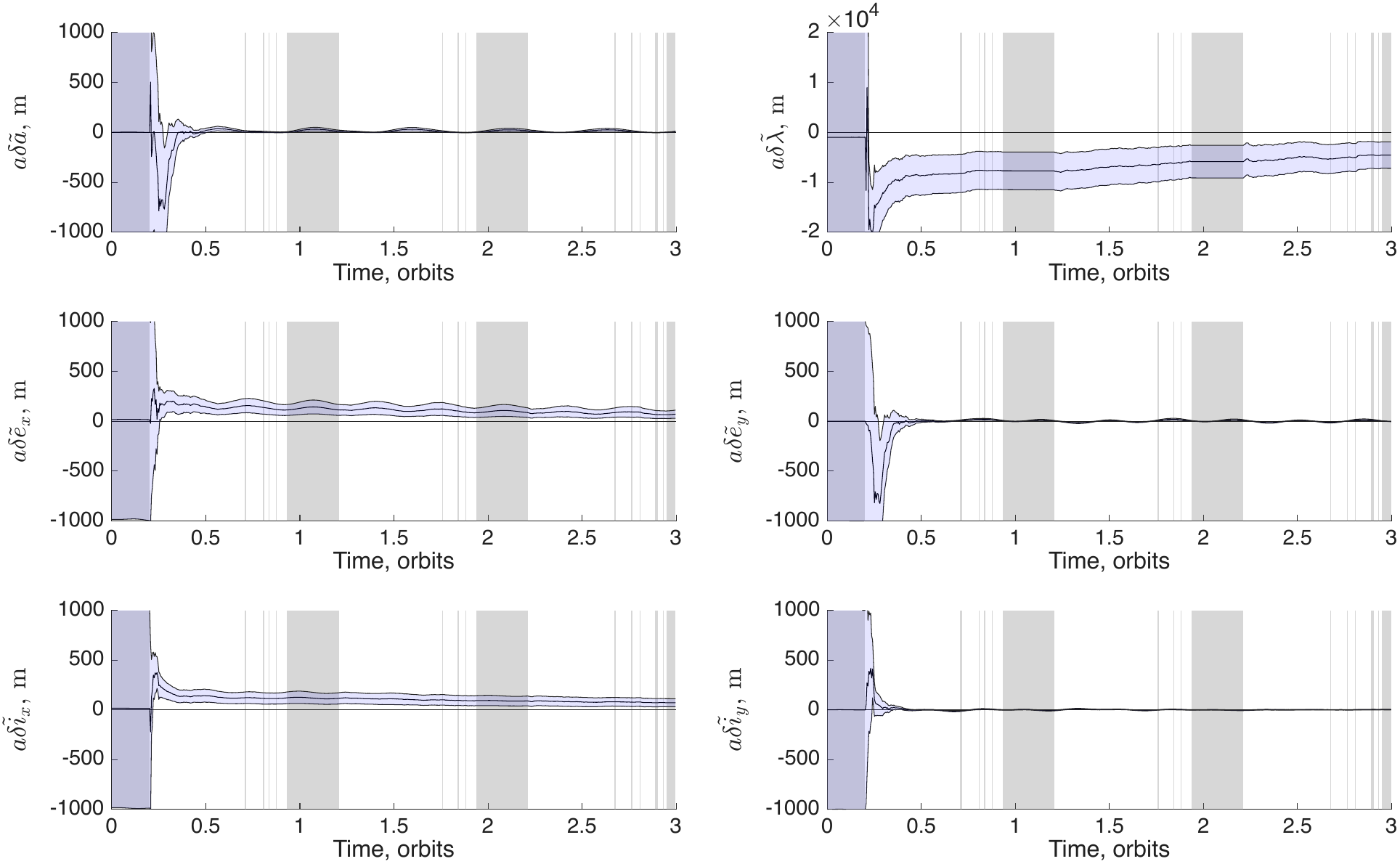}
     \caption{ROE estimation error of flat plate target without light curve measurements.}
     \label{fig:01convROE}
\end{figure}

Figure~\ref{fig:lcFlatPlateLEO} shows an example of the light curves for the flat plate target. The gray regions mark epochs with no available optical observation, in which the total observed flux vanishes and the apparent magnitude is undefined. The wide bands are Earth-shadow eclipses, which occur once per orbit, whereas the thin lines occur when the tumbling target presents no sunlit, observer-facing facet, that is, $\bm{n}_{i}^{T}\bm{s}\le 0$ or $\bm{n}_{i}^{T}\bm{v}\le 0$ for all facets. The thin lines are therefore not eclipses. Because they are governed by the attitude motion, they recur at the target's tumbling frequency, which is much higher than the orbital frequency, and hence appear many times within a single orbit.
Figure~\ref{fig:lcSensitivity} further illustrates how the light curve responds to the target attitude and to the optical properties over the sunlit arc of the first orbit. Varying the initial attitude completely reshapes the fast modulation, since the illumination history of every facet changes. Varying the diffuse reflectance shifts the curve almost uniformly by $-2.5\log_{10}$ of the reflectance ratio, and varying the specular reflectance changes the diffuse level through the $1-F_{0}$ factor together with the glint amplitudes. All of these dependencies are confined to the absolute level term of Eq.~\eqref{eq:mappsep}, whereas the range term $5\log_{10}r_{\rm obs}$ is common to every curve. This is why the recovered relative state tolerates uncertainty in the optical properties, while the attitude, which shapes the modulation, is supplied as a known input.
The apparent magnitude varies periodically with the target's tumbling motion, and sharp peaks correspond to geometrical conditions where the specular reflection is directed toward the observer.
Fig.~\ref{fig:01convROE} presents the ROE estimation error using only bearing angle measurements, where the tilde denotes the estimation error, i.e., the estimate minus the true value, $\tilde{(\cdot)}=\widehat{(\cdot)}-(\cdot)$.
The result clearly shows that the relative distance $\delta \lambda$ exhibits slow convergence due to the inherently weak observability of angles-only navigation.
This slow convergence of $\delta \lambda$ also affects the estimation of $\delta e_{x}$ and $\delta i_{x}$, since their observability is coupled through the relative geometry.

\begin{figure}[tb]
     \centering
     \includegraphics[width=12cm]{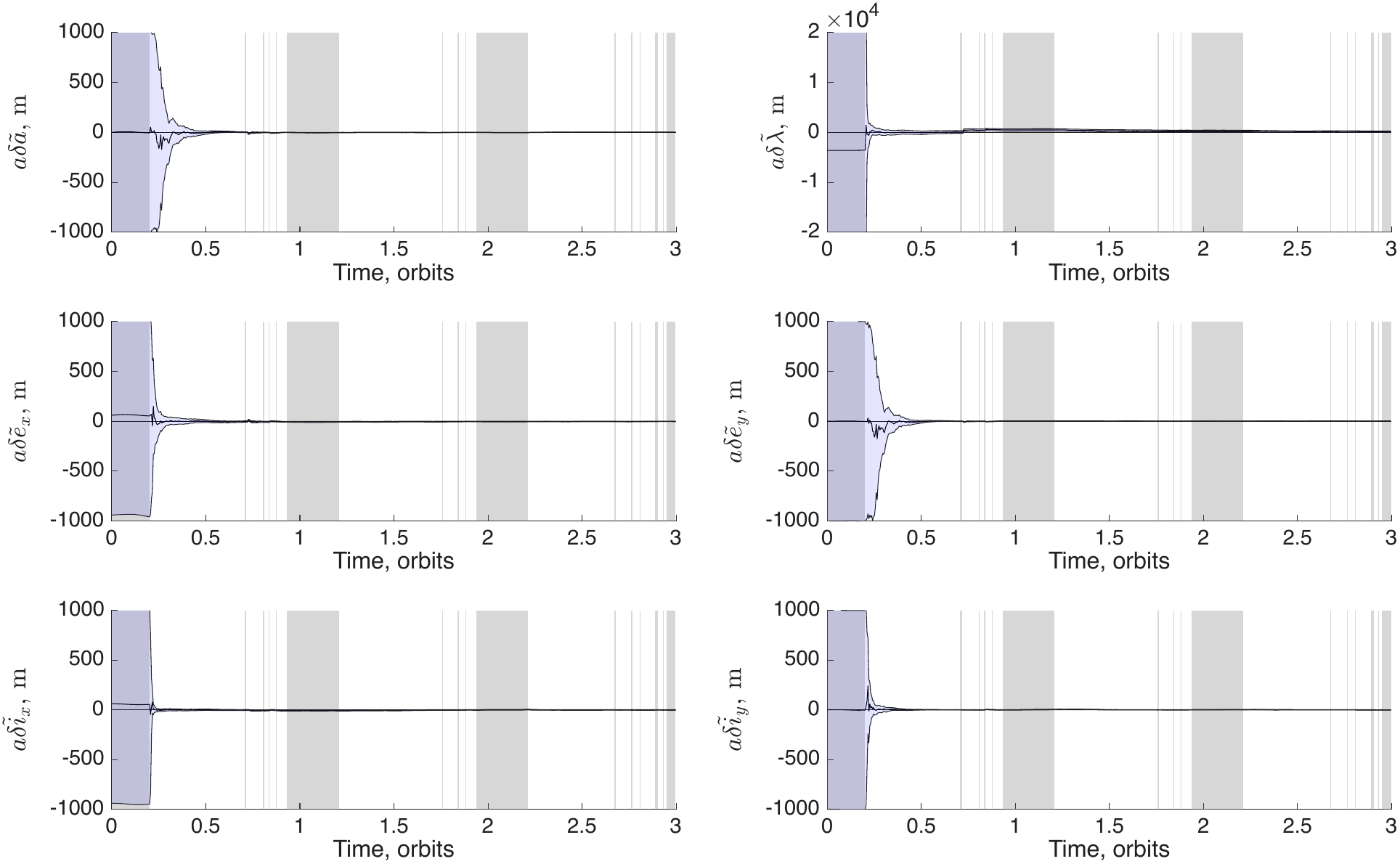}
     \caption{ROE estimation error of flat plate target with light curve measurements.}
     \label{fig:01attitudeGivenROE}
\end{figure}

In contrast, Fig.~\ref{fig:01attitudeGivenROE} shows the ROE estimation result when both bearing angles and light curves are used, assuming that the diffuse reflectance $\rho_d$ is known a priori.
The relative distance $a\delta\lambda$ converges rapidly to the true value within approximately one orbital period when observations are available, demonstrating a significant improvement in convergence compared to the previous case.
The other ROE components are also estimated with fast convergence, confirming that the light curve measurements effectively enhance the observability of the relative state.

\begin{figure}[tb]
     \centering
     \includegraphics[width=12cm]{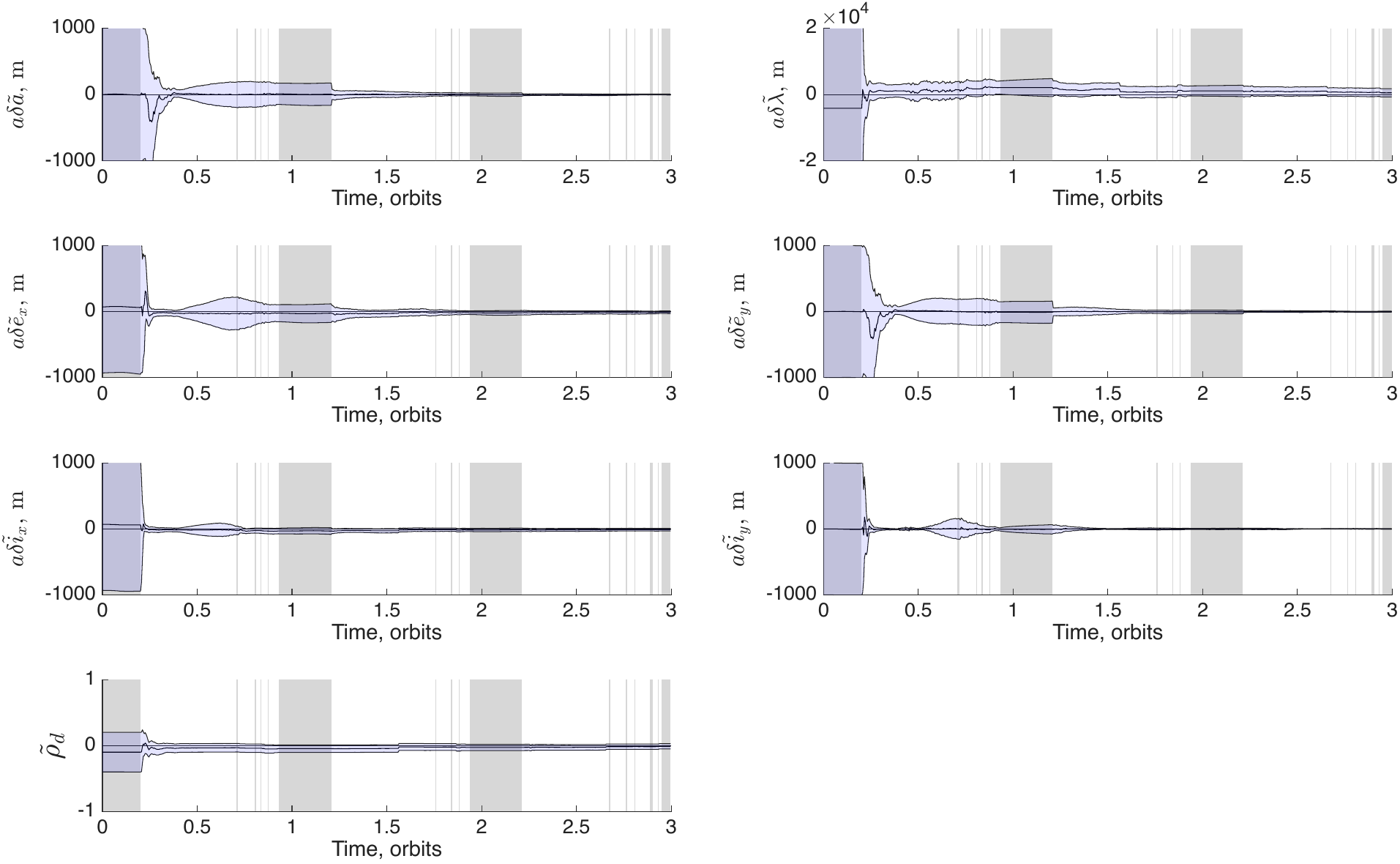}
     \caption{ROE and optical property estimation errors of flat plate target with light curve measurements.}
     \label{fig:flatPlateLEOsingle}
\end{figure}

Figure~\ref{fig:flatPlateLEOsingle} presents the results of simultaneous estimation of the ROE and the diffuse reflectance $\rho_d$ for the flat plate target.
Compared to the case where $\rho_d$ is known in Fig.~\ref{fig:01attitudeGivenROE}, the convergence rate is slower because the estimator must resolve the coupling between the relative distance and the optical property.
Nevertheless, the ROE estimation converges successfully within approximately two orbital periods, indicating that the augmented state approach is effective for practical scenarios where the target's optical properties are uncertain.
As shown in $\tilde{\rho}_d$ of Fig.~\ref{fig:flatPlateLEOsingle}, the diffuse reflectance estimate converges to within a few percent of its true value and then remains nearly constant.
The small residual variations, on the order of a few percent of $\rho_d$, reflect the filter response to the time-varying light curve and the eclipse-induced measurement gaps rather than any change in the true reflectance. Thus, treating $\rho_d$ as a constant remains consistent with the estimated behavior.

\begin{figure}[tb]
     \centering
     \includegraphics[width=12cm]{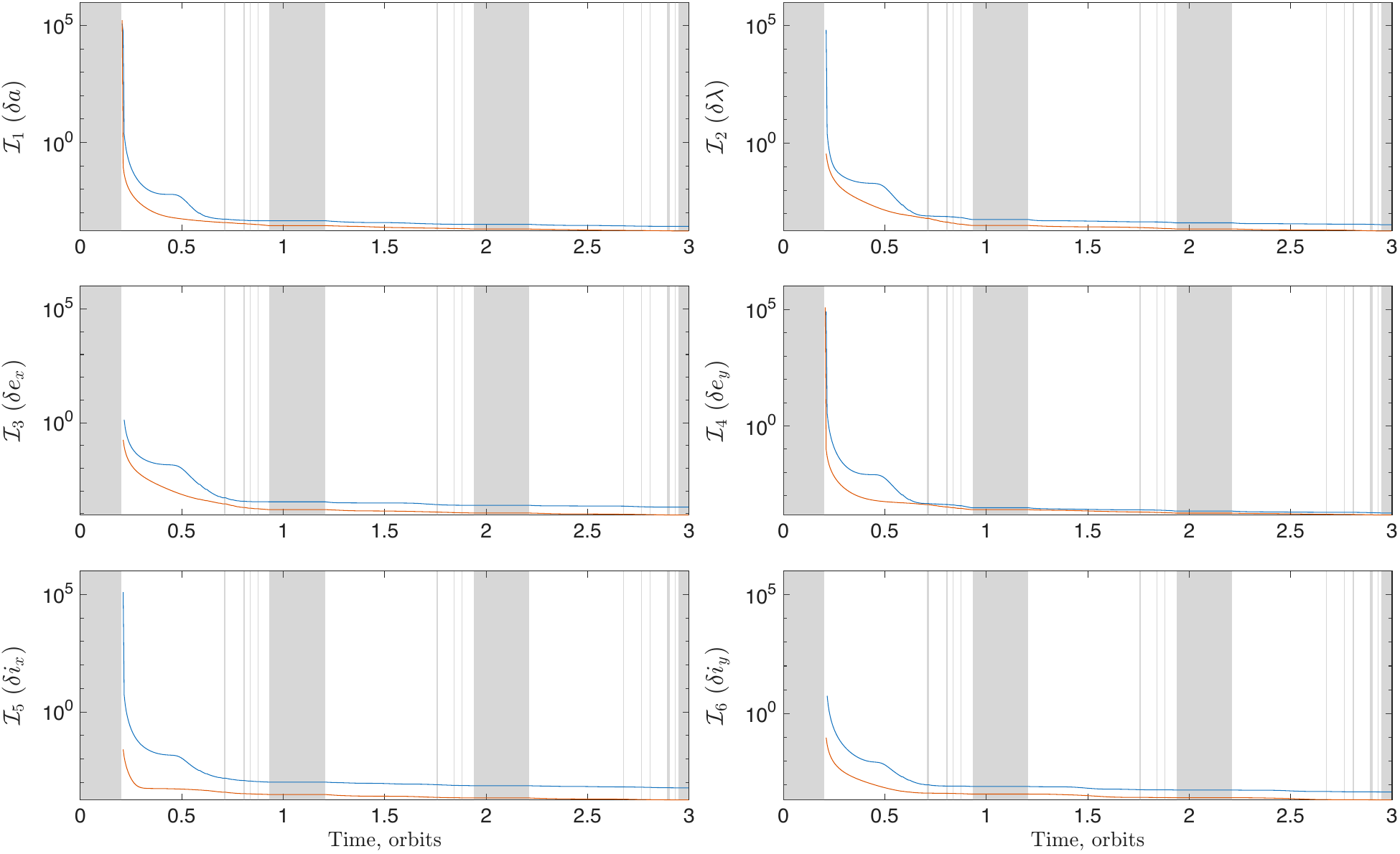}
     \caption{Observability analysis of ROE of flat plate target. The Fisher information matrix is formed for the relative state only, so the index is a known-optics bound.}
     \label{fig:01flatPlateLEOobs}
\end{figure}

Figure~\ref{fig:01flatPlateLEOobs} presents the observability analysis based on the Fisher information matrix for the flat plate target.
The blue line represents the observability index using bearing angle measurements only, while the red line shows the index using both bearing angles and light curves.
The observability index with light curves decreases much more rapidly, indicating higher observability.
This result is consistent with the estimation results. That is, the faster decrease of the observability index corresponds to the faster convergence observed in Fig.~\ref{fig:01attitudeGivenROE}.

\begin{figure}[tb]
     \centering
     \includegraphics[width=12cm]{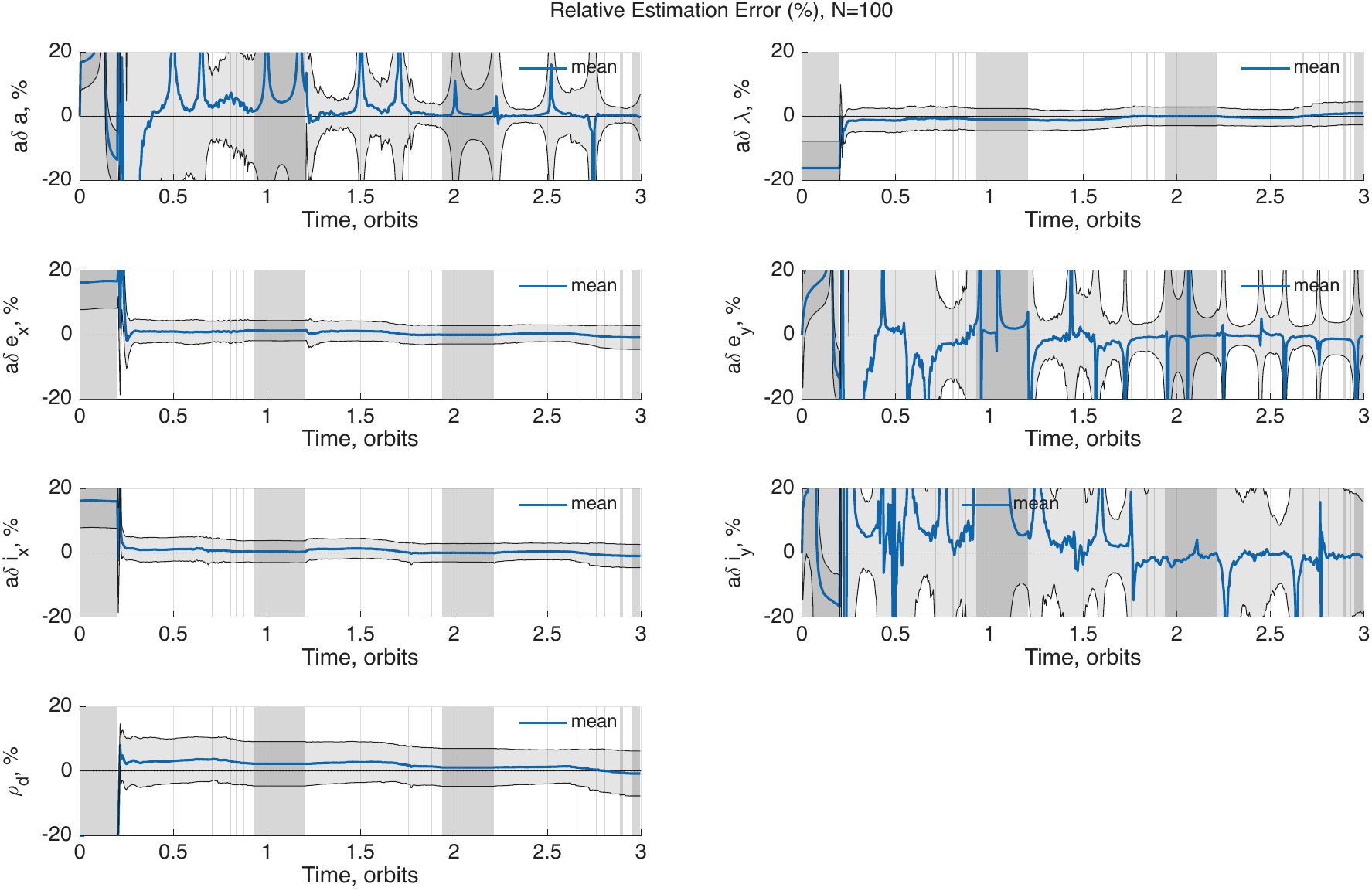}
     \caption{Relative errors of relative distance and diffuse reflectance for flat plate target.}
     \label{fig:flatPlateLEOrel}
\end{figure}

Table~\ref{tab:01rmse} summarizes the root-mean-square error (RMSE) and standard deviation of the ROE estimates over 100 Monte Carlo runs with simultaneous estimation of $\rho_d$.
All 100 runs converge to the true relative state from random initializations with up to 30\% initial error, indicating that the basin of attraction of the true solution covers the tested initialization domain and that convergence to a spurious $\delta\lambda$ does not occur. This is consistent with the observability analysis in Fig.~\ref{fig:01flatPlateLEOobs}, where the light-curve term makes $\delta\lambda$ well-conditioned rather than leaving the flat valley of the angles-only case.
Figure~\ref{fig:flatPlateLEOrel} shows the relative errors of the relative distance and diffuse reflectance for the converged runs, verifying that the estimation achieves a relative error of a few percent even when the diffuse reflectance is uncertain.

The benefit of augmenting the state with $\rho_d$, discussed in Section~\ref{subsec:ukf}, is quantified in Table~\ref{tab:augComp}. Three treatments of the diffuse reflectance over the same 100 Monte Carlo runs with identical noise realizations and initializations are compared: $\rho_d$ fixed at its true value, $\rho_d$ fixed at a wrong value, and $\rho_d$ estimated online from different initial guesses. All cases converge in 100 of 100 runs, but they differ sharply in accuracy. With $\rho_d$ fixed at the true value, the estimate is essentially unbiased. This is the known-optics best case. With $\rho_d$ fixed at a wrong value, the recovered relative orbit carries a systematic bias along the line-of-sight-preserving scaling direction: an assumed reflectance $\rho_{\rm a}$ displaces the brightness level by $-2.5\log_{10}(\rho_{\rm a}/\rho_{d})$, which the range term of Eq.~\eqref{eq:mappsep} absorbs as a scale factor of approximately $\sqrt{\rho_{\rm a}/\rho_{d}}$ on the entire relative orbit. The measured biases of $a\delta\lambda$, $+3.13$~km for $\rho_{\rm a}=0.4$ and $+6.58$~km for $\rho_{\rm a}=0.3$, agree with this prediction ($+3.17$ and $+6.76$~km) to within a few percent. Estimating $\rho_d$ online removes this systematic error. For initial guesses spanning $0.2$ to $0.8$, i.e., up to $60\%$ prior error, the residual bias stays within $\pm0.8$~km, comparable to the run-to-run scatter and roughly fifteen times smaller than the bias that the same prior error would induce with a fixed $\rho_d$. The reflectance itself is recovered with an RMSE of $0.033$--$0.039$ irrespective of the initial guess. The box-wing target behaves identically (a $+3.13$~km bias for fixed $\rho_{\rm a}=0.4$ and residual biases within $\pm0.8$~km when estimated). The augmentation thus converts an uncorrectable systematic range-scale bias into a small estimation variance at the cost of a single additional state.

The last row of Table~\ref{tab:augComp} examines whether this separation survives when the attitude-driven brightness modulation is removed. The target attitude is held inertially fixed, while $\rho_d$ is estimated from the initial guess $0.4$. All 100 runs converge with a bias of $-0.06\pm1.16$~km, i.e., no systematic error, because the two remaining brightness contributions are still distinguishable. A wrong constant reflectance offsets the absolute level of Eq.~\eqref{eq:mappsep} uniformly over the pass, whereas the range term $5\log_{10}r_{\rm obs}$ retains its time variation from the relative orbital geometry even without attitude motion. The reduced brightness diversity does weaken the separation, which appears as a larger scatter ($1.644$~km RMSE versus $1.091$~km for the tumbling baseline, and $0.053$ versus $0.035$ for $\rho_{d}$), but the mismodeled brightness is absorbed by the reflectance state as designed rather than being misattributed to the range.

\begin{figure}[tb]
     \centering
     \includegraphics[width=9cm]{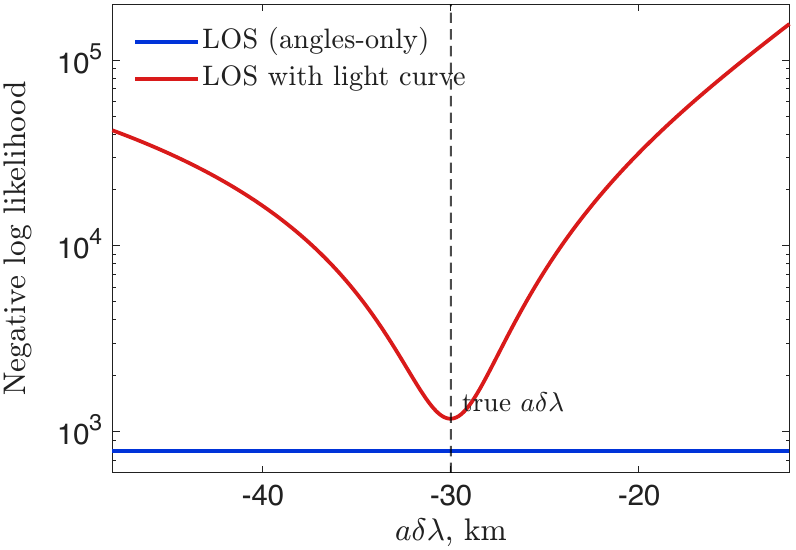}
     \caption{Negative log-likelihood (batch innovation cost) along the line-of-sight-preserving scaling direction of the relative state, parametrized by the relative mean longitude $a\delta\lambda$.  The dashed line marks the true $a\delta\lambda$.}
     \label{fig:costLambda}
\end{figure}

To make explicit why the light curve excludes a spuriously scaled relative orbit, Fig.~\ref{fig:costLambda} shows the batch innovation cost (negative log-likelihood) evaluated along the line-of-sight-preserving scaling direction of the relative state, parametrized by the relative mean longitude $a\delta\lambda$. Scaling the entire relative orbit leaves the bearing directions unchanged, so the angles-only cost is flat: every $a\delta\lambda$ along this direction fits the bearings equally well, which is the range ambiguity of angles-only navigation. Adding the light curve introduces the range term $5\log_{10}r_{\rm obs}$ of Eq.~\eqref{eq:mappsep}, which turns the cost into a single sharp minimum at the true value. The light curve therefore removes the flat valley and excludes convergence to a wrong $\delta\lambda$, in agreement with the Fisher-information result of Fig.~\ref{fig:01flatPlateLEOobs}.

\begin{table}[tb]
     \begin{center}
          \caption{RMSE and standard deviation of ROE estimates for 100 Monte Carlo runs (flat plate target).}\label{tab:01rmse}
          \begin{tabular}{lccc}
               \hline \hline
               State             & Unit & RMSE       & Standard deviation \\ \hline
               $a\delta a$       & m    & $2.029$    & $2.036$            \\
               $a\delta \lambda$ & m    & $1090.684$ & $1063.692$         \\
               $a\delta e_{x}$   & m    & $17.655$   & $17.288$           \\
               $a\delta e_{y}$   & m    & $1.623$    & $1.630$            \\
               $a\delta i_{x}$   & m    & $17.103$   & $16.611$           \\
               $a\delta i_{y}$   & m    & $2.290$    & $2.296$            \\
               $\rho_{d}$        & --   & $0.035$    & $0.035$            \\
               \hline \hline
          \end{tabular}
     \end{center}
\end{table}

\begin{table}[tb]
     \begin{center}
          \caption{Effect of the $\rho_{d}$ treatment on the relative orbit estimate for 100 Monte Carlo runs. The bias of $a\delta\lambda$ over the final orbital period is given as mean $\pm$ run-to-run standard deviation and as a percentage of the true value ($30$~km). The last row holds the target attitude inertially fixed.}\label{tab:augComp}
          \setlength{\tabcolsep}{3pt}
          \begin{tabular}{llccccc}
               \hline \hline
               $\rho_{d}$ treatment &                     & Bias of $a\delta\lambda$~km & Bias~\% & Predicted bias~km & RMSE of $a\delta\lambda$~km & RMSE of $\rho_{d}$ \\ \hline
               Fixed                & at true value $0.5$ & $+0.09\pm0.13$              & $+0.3$  & $0$               & $0.204$                     & --                 \\
               Fixed                & at $0.4$            & $+3.13\pm0.12$              & $+10.4$ & $+3.17$           & $3.173$                     & --                 \\
               Fixed                & at $0.3$            & $+6.58\pm0.13$              & $+21.9$ & $+6.76$           & $6.618$                     & --                 \\
               Estimated            & initial guess $0.2$ & $+0.70\pm0.75$              & $+2.3$  & $0$               & $1.284$                     & $0.039$            \\
               Estimated            & initial guess $0.35$& $-0.02\pm0.81$              & $-0.1$  & $0$               & $1.089$                     & $0.035$            \\
               Estimated            & initial guess $0.4$ & $+0.01\pm0.82$              & $+0.0$  & $0$               & $1.091$                     & $0.035$            \\
               Estimated            & initial guess $0.5$ & $-0.04\pm0.79$              & $-0.1$  & $0$               & $1.038$                     & $0.034$            \\
               Estimated            & initial guess $0.65$& $-0.22\pm0.76$              & $-0.7$  & $0$               & $0.988$                     & $0.033$            \\
               Estimated            & initial guess $0.8$ & $-0.52\pm0.74$              & $-1.7$  & $0$               & $1.005$                     & $0.035$            \\
               Estimated            & constant attitude, guess $0.4$ & $-0.06\pm1.16$   & $-0.2$  & $0$               & $1.644$                     & $0.053$            \\
               \hline \hline
          \end{tabular}
     \end{center}
\end{table}

\subsection{Box-wing Target}
\begin{figure}[tb]
     \centering
     \includegraphics[width=11cm]{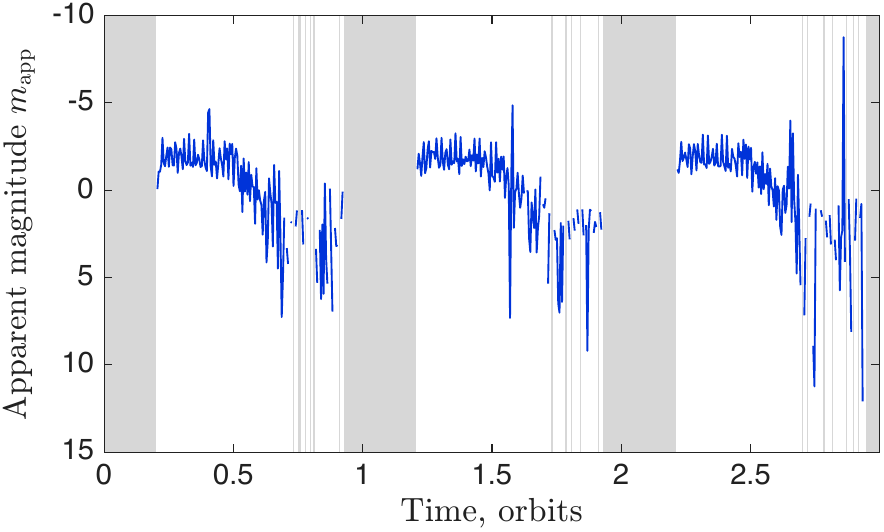}
     \caption{Light curves of box-wing target in LEO.}
     \label{fig:2aLC}
\end{figure}

To examine the applicability of the proposed approach to more realistic target geometries, the box-wing satellite is considered as a second test case.
Figure~\ref{fig:2aLC} presents the light curves of the box-wing target in LEO.
Although the orbital conditions are identical to the flat plate case, the light curve exhibits a distinctly different time history due to the complex geometry of the box-wing object.
In particular, the non-convex shape introduces self-shadowing, which causes abrupt changes in the apparent magnitude when certain facets enter or exit the shadow of other structural components.

\begin{figure}[tb]
     \centering
     \includegraphics[width = 12cm]{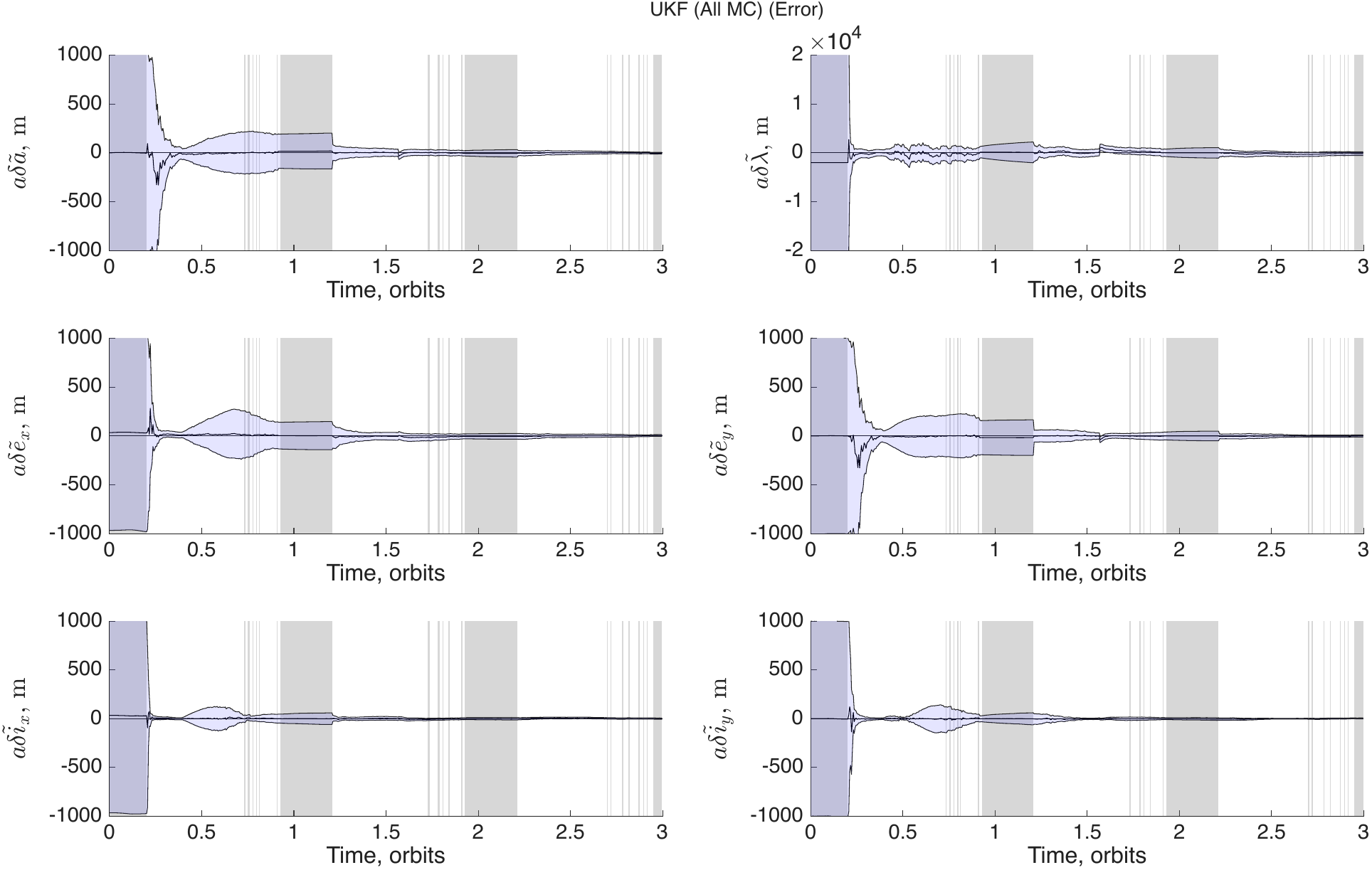}
     \caption{ROE estimation error of box-wing target with light curve measurements.}
     \label{fig:2aROEgiven}
\end{figure}

Figure~\ref{fig:2aROEgiven} shows the ROE estimation error when both bearing angles and light curves are used, assuming that the diffuse reflectance is known.
Despite the complex geometry, the relative distance $\delta \lambda$ converges rapidly, confirming that the light curve measurements provide sufficient observability enhancement regardless of the target shape.
Note that since the initial estimate has a large error up to 30\%, the initial estimate of $a\delta \lambda$ is outside the plotted range in Fig.~\ref{fig:2aROEgiven}. However, the error decreases rapidly once observations become available.
The observability analysis for the box-wing target is omitted because the trend of the observability index is qualitatively similar to that of the flat plate case.

\begin{figure}[tb]
     \centering
     \includegraphics[width = 12cm]{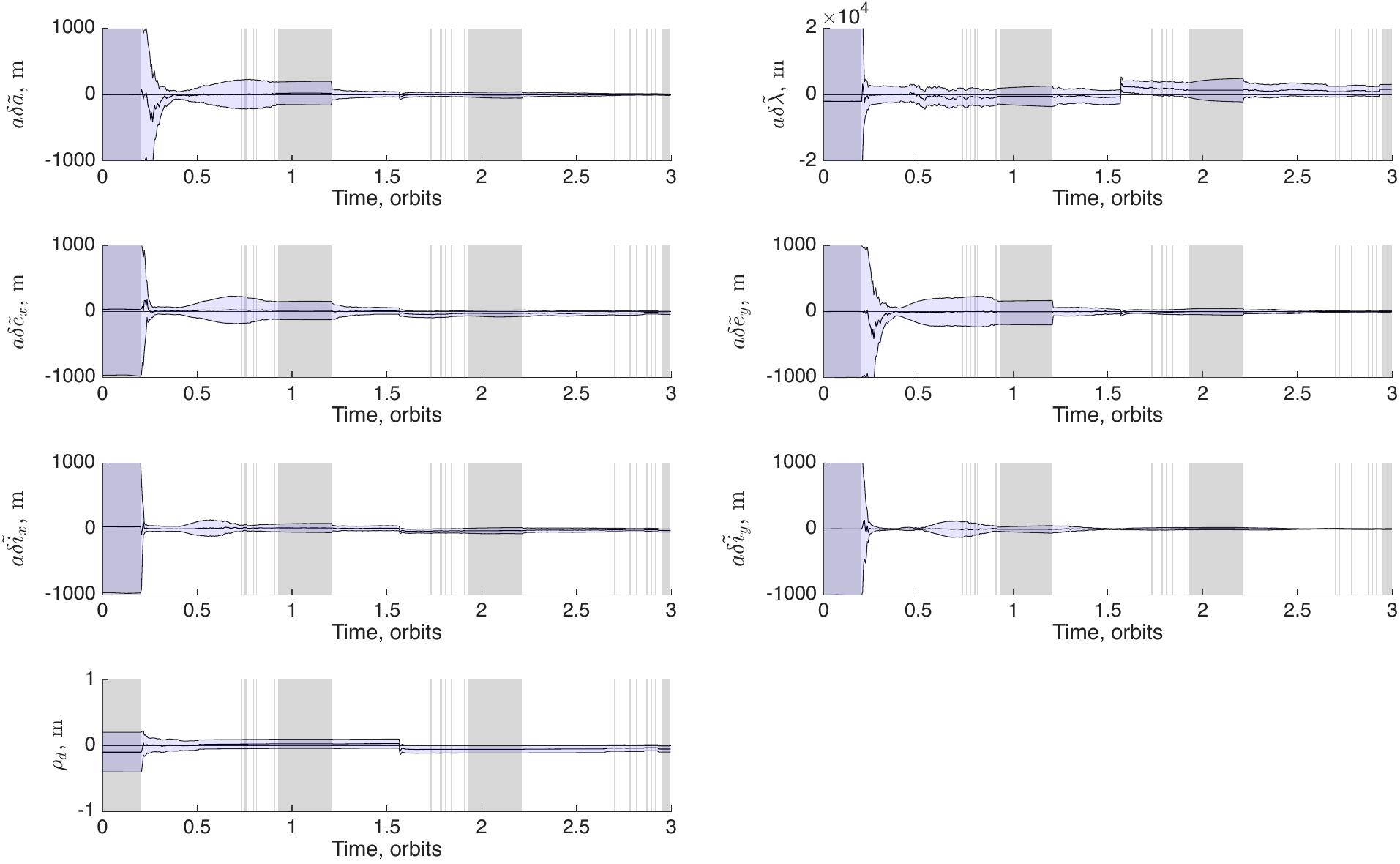}
     \caption{ROE and optical property estimation errors of box-wing target with light curve measurements.}
     \label{fig:boxWingLEOsingle}
\end{figure}

\begin{figure}[tb]
     \centering
     \includegraphics[width=12cm]{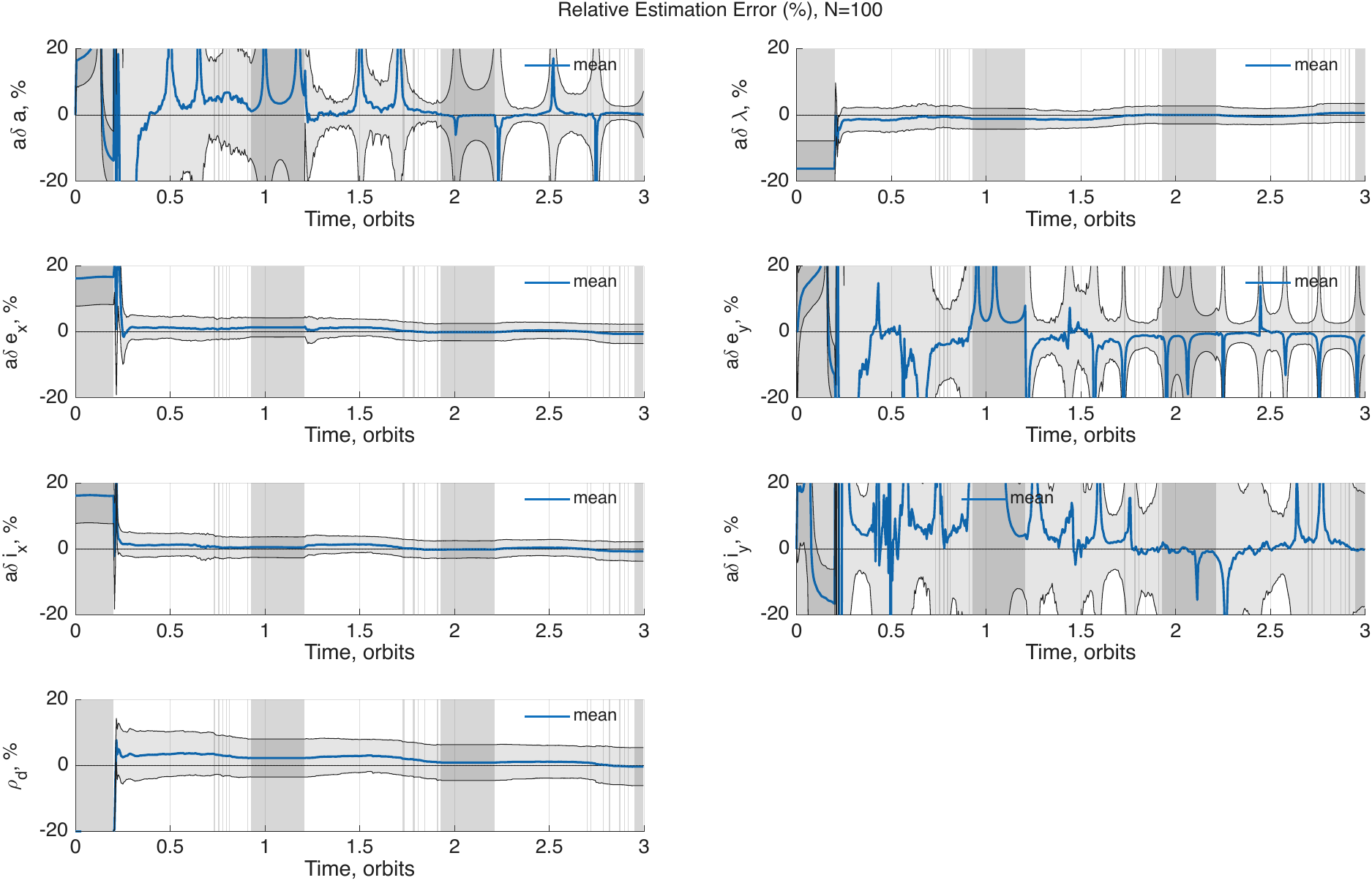}
     \caption{Relative errors of relative distance and diffuse reflectance for box-wing target.}
     \label{fig:boxWingLEOrel}
\end{figure}

Table~\ref{tab:02rmse} summarizes the RMSE and standard deviation of the ROE estimates over 100 Monte Carlo runs with simultaneous estimation of $\rho_d$.
As in the flat plate case, all 100 runs converge.
The RMSE values are also of the same order of magnitude as those of the flat plate target, suggesting that the estimation performance is not significantly degraded by the increased geometric complexity or the presence of self-shadowing.
Figure~\ref{fig:boxWingLEOsingle} shows the estimate errors of ROE and diffuse reflectance, and Figure~\ref{fig:boxWingLEOrel} shows the relative errors of the relative distance and diffuse reflectance for the converged runs, confirming that the estimation achieves relative errors of a few percent.
These results demonstrate that the proposed estimation scheme combining bearing angles and light curves is effective for both convex and non-convex target geometries.

\begin{table}[tb]
     \begin{center}
          \caption{RMSE and standard deviation of ROE estimates for 100 Monte Carlo runs (box-wing target).}\label{tab:02rmse}
          \begin{tabular}{lccc}
               \hline \hline
               State             & Unit & RMSE       & Standard deviation \\ \hline
               $a\delta a$       & m    & $1.914$    & $1.923$            \\
               $a\delta \lambda$ & m    & $858.096$  & $841.208$          \\
               $a\delta e_{x}$   & m    & $13.996$   & $13.775$           \\
               $a\delta e_{y}$   & m    & $1.637$    & $1.626$            \\
               $a\delta i_{x}$   & m    & $13.923$   & $13.594$           \\
               $a\delta i_{y}$   & m    & $2.249$    & $2.260$            \\
               $\rho_{d}$        & --   & $0.029$    & $0.029$            \\
               \hline \hline
          \end{tabular}
     \end{center}
\end{table}

\subsection{Sensitivity Analysis}\label{subsec:sens}
To assess the dependence of the preceding results on the assumed scenario conditions, the Monte Carlo simulation is repeated with one condition varied at a time: the initial relative orbit, the illumination conditions, and the fidelity of the target shape model. Each variation uses 100 runs with the same randomization design as before and online estimation of $\rho_{d}$ from the initial guess $0.4$. Table~\ref{tab:sens} summarizes the results.

Doubling the initial relative distance to $a\delta\lambda_{0}=-60$~km preserves convergence in all runs, with the estimation error growing roughly in proportion to the range, reflecting the weaker bearing geometry and the dimmer target. Reducing the distance to $-10$~km yields the most accurate results of all cases (RMSE $0.25$~km) provided the initial standard deviation of $a\delta\lambda$ is scaled to the geometry. If instead the baseline value of $50$~km, five times the actual separation, is retained, the filter drifts along the level--range direction. A scaled relative orbit $r\to c\,r$ with reflectance $c^{2}\rho_{d}$ reproduces the same bearings and magnitudes. An excessively wide prior on the range lets the estimate migrate along this family until the physical bound $\rho_{d}\le 1$ is reached, and only 1 of 100 runs then remains near the truth. This sensitivity arises because the level--range family is the weakly observable direction of the fused problem. Constraining $\rho_{d}$ to its physical range and choosing the initial covariance commensurate with the expected separation are the corresponding safeguards.

Shifting the epoch from January~1 to April~1 and July~1, 2021 changes the Sun direction, the eclipse pattern, and the observation availability. All runs converge with accuracy equal to or better than the baseline, indicating that the conclusions are not tied to the particular illumination geometry of the baseline epoch.

Anticipating that any box-wing model is a simplification of a real spacecraft, the truth data are regenerated with a finer box-wing model (2124 facets versus 472), whose apparent magnitude differs from that of the coarse model by $1.4$~mag RMS at identical states, while the filter retains the coarse model: an intentionally severe geometric-mismatch test. For this test self-shadowing is disabled on both models so that the comparison isolates the effect of facet resolution. Enabling it would enlarge the discrepancy, but since a magnitude error of any origin maps onto the same level--range decomposition, it would not alter the conclusion. No abrupt failures occur, and all 100 runs converge smoothly. The systematic photometric model error is split by the estimator between the effective reflectance, which settles at $\rho_{d}\approx 0.27$, and a range-scale bias of $-11$~km, consistent with the level--range structure of Eq.~\eqref{eq:mappsep}. The part of the level error not absorbed by $\rho_{d}$, about $0.7$~mag, maps onto the scale factor $10^{0.7/5}\approx 1.4$. Shape simplification therefore does not destabilize the filter; it bounds the absolute range accuracy through the photometric fidelity of the shape model, with the residual error confined to the scale direction.

Across all cases the residual systematic error lies on the level--range family $(r,\rho_{d})\to(c\,r,\,c^{2}\rho_{d})$. For example, at $-60$~km the converged reflectance $0.47$ and the $+2.1$~km bias satisfy $\rho_{d}\approx c^{2}\rho_{d,\rm true}$ with $c=0.97$ the realized scale factor. The fused estimator thus degrades gracefully along this single predictable direction rather than failing abruptly, which is the practically relevant property for deploying the method with imperfect prior knowledge.

\begin{table}[tb]
     \begin{center}
          \caption{Sensitivity of the estimation performance to the scenario conditions. The bias of $a\delta\lambda$ over the final orbital period is given also as a percentage of the true value of each case; the RMSE is taken at the final time over the converged runs.}\label{tab:sens}
          \begin{tabular}{lccccc}
               \hline \hline
               Case                                                & Converged & Bias of $a\delta\lambda$~km & Bias~\% & RMSE of $a\delta\lambda$~km & Final $\rho_{d}$ \\ \hline
               Baseline ($a\delta\lambda_{0}=-30$~km, Jan.~1)      & $100/100$ & $+0.01\pm0.82$              & $+0.0$  & $1.091$                     & $0.50$           \\
               $a\delta\lambda_{0}=-10$~km ($\sigma_{a\delta\lambda}=15$~km) & $100/100$ & $-0.08\pm0.26$    & $-0.8$  & $0.253$                     & $0.51$           \\
               $a\delta\lambda_{0}=-60$~km                         & $100/100$ & $+2.08\pm2.98$              & $+3.5$  & $3.698$                     & $0.47$           \\
               Epoch Apr.~1 (valid fraction $0.71$)                & $100/100$ & $-0.01\pm0.51$              & $-0.0$  & $0.508$                     & $0.50$           \\
               Epoch Jul.~1 (valid fraction $0.66$)                & $100/100$ & $+0.03\pm0.55$              & $+0.1$  & $0.574$                     & $0.50$           \\
               Box-wing, fine-model truth                          & $100/100$ & $-11.03\pm1.92$             & $-36.8$ & $11.924$                    & $0.27$           \\
               \hline \hline
          \end{tabular}
     \end{center}
\end{table}

\section{Conclusions}
This paper deals with relative orbit estimation using bearing angles and light curves.
Conventional angles-only navigation suffers from the weak observability of relative distance.
This paper employs the adaptive unscented Kalman filter for estimating ROE.
When light curve measurements are used with the bearing angles, the relative distance component of the ROE converges quickly, indicating the effectiveness of including light curve measurements.
Furthermore, the ROE estimate is robust even if the initial estimate has a 30\% error, indicating that precise initial orbit knowledge is not required when light curves are used.

% \appendix
% \section{My Appendix}

\printcredits

% Loading bibliography style file
\bibliographystyle{unsrtnat}
%\bibliographystyle{cas-model2-names}

\bibliography{roeEst}


% Author biography without author photo.
% Author biography. Author biography. Author biography.
% Author biography. Author biography. Author biography.
% Author biography. Author biography. Author biography.
% Author biography. Author biography. Author biography.
% Author biography. Author biography. Author biography.
% Author biography. Author biography. Author biography.
% Author biography. Author biography. Author biography.
% Author biography. Author biography. Author biography.
% Author biography. Author biography. Author biography.

% % \bio{figs/cas-pic1}
% Author biography with author photo.
% Author biography. Author biography. Author biography.
% Author biography. Author biography. Author biography.
% Author biography. Author biography. Author biography.
% Author biography. Author biography. Author biography.
% Author biography. Author biography. Author biography.
% Author biography. Author biography. Author biography.
% Author biography. Author biography. Author biography.
% Author biography. Author biography. Author biography.
% Author biography. Author biography. Author biography.

% \vskip3pc

% % \bio{figs/cas-pic1}
% Author biography with author photo.
% Author biography. Author biography. Author biography.
% Author biography. Author biography. Author biography.
% Author biography. Author biography. Author biography.
% Author biography. Author biography. Author biography.

\end{document}